\RequirePackage{fix-cm}
\documentclass[aps,pra,reprint,superscriptaddress]{revtex4-2}

\RequirePackage{graphicx}
\usepackage{graphicx}
\usepackage{dcolumn}
\usepackage{bm}
\usepackage{lipsum}
\usepackage{subfigure}

\usepackage{tikz}
\usepackage{ctable}
\usetikzlibrary{arrows.meta}
\usepackage{pgfplots}
\usepackage{amsmath}
\usepgfplotslibrary{colormaps}
\pgfplotsset{compat=1.8}
\usetikzlibrary{pgfplots.groupplots,pgfplots.fillbetween}

\usepackage[colorlinks=true,linkcolor=black,citecolor=blue,urlcolor=blue,]{hyperref}

\def \rthe{$\rm {^{3}H/^{3}He}$}
\def \esym{$E_{\rm sym}(\rho)$}

\def\rthe{$\rm {^{3}H/^{3}He}$}
\def \esym{$E_{\rm sym}(\rho)$}

\begin{document}

\title{Studies of nuclear equation of state with the HIRFL-CSR external-target experiment}
\author{Dong Guo} 
\email[]{guodong19@mails.tsinghua.edu.cn}
\affiliation{Department of Physics, Tsinghua University, Beijing 100084, China}
\author{Xionghong He} 
\affiliation{Institute of Modern Physics, Chinese Academy of Science, Lanzhou 730000, China}
\author{Pengcheng Li} 
\affiliation{School of Science, Huzhou University, Huzhou 313000, China}
\author{Zhi Qin} 
\affiliation{Department of Physics, Tsinghua University, Beijing 100084, China}
\author{Chenlu Hu} 
\affiliation{Institute of Modern Physics, Chinese Academy of Science, Lanzhou 730000, China}
\author{Botan Wang} 
\affiliation{Department of Engineering Physics, Tsinghua University, Beijing 100084, China}
\author{Yingjie Zhou} 
\affiliation{Department of Modern Physics, University of Science and Technology of China, Hefei 230026, China}
\author{Kun Zheng} 
\affiliation{Department of Physics, Tsinghua University, Beijing 100084, China}
\author{Yapeng Zhang} 
\affiliation{Institute of Modern Physics, Chinese Academy of Science, Lanzhou 730000, China}
\author{Xianglun Wei} 
\affiliation{Institute of Modern Physics, Chinese Academy of Science, Lanzhou 730000, China}
\author{Herun Yang} 
\affiliation{Institute of Modern Physics, Chinese Academy of Science, Lanzhou 730000, China}
\author{Dongdong Hu} 
\affiliation{Department of Modern Physics, University of Science and Technology of China, Hefei 230026, China}
\author{Ming Shao} 
\affiliation{Department of Modern Physics, University of Science and Technology of China, Hefei 230026, China}
\author{Limin Duan} 
\affiliation{Institute of Modern Physics, Chinese Academy of Science, Lanzhou 730000, China}
\author{Yuhong Yu} 
\affiliation{Institute of Modern Physics, Chinese Academy of Science, Lanzhou 730000, China}
\author{Zhiyu Sun} 
\affiliation{Institute of Modern Physics, Chinese Academy of Science, Lanzhou 730000, China}
\author{Yongjia Wang} 
\affiliation{School of Science, Huzhou University, Huzhou 313000, China}
\author{Qingfeng Li} 
\affiliation{School of Science, Huzhou University, Huzhou 313000, China}
\affiliation{Institute of Modern Physics, Chinese Academy of Science, Lanzhou 730000, China}
\author{Zhigang Xiao} 
\email[]{xiaozg@mail.tsinghua.edu.cn}
\affiliation{Department of Physics, Tsinghua University, Beijing 100084, China}

\begin{abstract} 
The HIRFL-CSR external-target experiment (CEE) under construction is expected to provide novel opportunities to the studies of the thermodynamic properties, namely the equation of state of nuclear matter (nEOS) with heavy ion collisions at a few hundreds MeV/u  beam energies. Based on Geant 4 packages, the fast simulations of the detector responses to the collision events generated using transport model are conducted. The overall  performance of CEE, including spatial resolution of hits,  momentum resolution of tracks and particle identification ability has been investigated. Various observables proposed to probe the nEOS, such as the production of light clusters, $\rm t/^3He$ yield ratio, the radial flow, $\pi^{-}/\pi^{+}$ yield ratio and the neutral  kaon yields, have been reconstructed. The  feasibility of studying nEOS beyond the saturation density via the aforementioned  observables to be measured with CEE has been demonstrated.


\end{abstract}

\maketitle


\section{Introduction}\label{sec. I}

\begin{figure}[!htb] 
\centering
\hspace{-0.5cm}\includegraphics[width=0.42\textwidth]{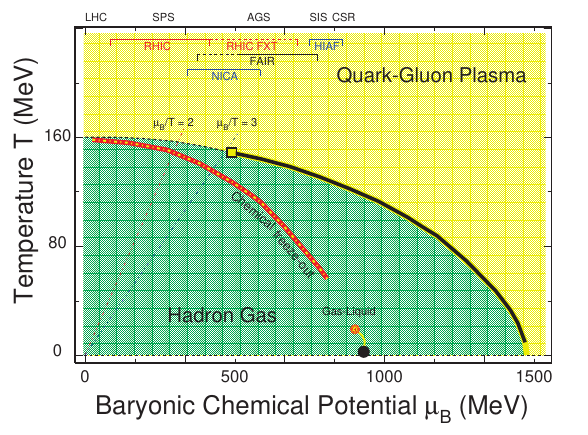}
\caption{(Color online) QCD phase structure diagram of temperature and baryon chemical potential description ~\cite{Luo2017Search}. The red line in the figure is the chemical freeze-out curve of common nuclear substances obtained by experiments. The black square points and black solid lines represent the expected first-order phase boundaries and critical points. The red circle represents the CEP at which the nuclear material becomes gas-liquid phase. The study areas of RHIC beam energy scanning and fixed target (USA), NICA (Russia), FAIR (Germany), and HIAF (China) are shown at the top of the figure.}
\label{QCD phase diagram}
\end{figure}

The phase diagram  of nuclear matter is of significance in understanding the properties of the strong interaction described by Quantum Chromodynamics (QCD) and the evolution of the universe. With the relativistic heavy ion experiments, the various experimental signals have revealed that  strongly interacting  matter may undergo the transition from the hadron phase to the quark-gluon-plasma (QGP) phase, for which the transition at low baryon density and high temperature is crossover ~\cite{Arsene2005Quark, BACK2005PHOBOS, Braun2007QGP}. FIG.\ref{QCD phase diagram} shows the QCD phase diagram of current understanding ~\cite{Luo2017Search}. Despite of great progress along this direction,  a few important questions remain unsolved concerning the QCD phase diagram. For instance, what are the thermodynamic properties of QCD and where is the phase boundary between QGP and hadronic matter at high net-baryon density? Does the critical end point (CEP) exist and where is the location? To shed light on these questions,  it is of great significance to explore further the QCD phase diagram with more systematic studies.

\begin{figure}[htb]
\centering
\hspace{-0.5cm}\includegraphics[width=0.42\textwidth]{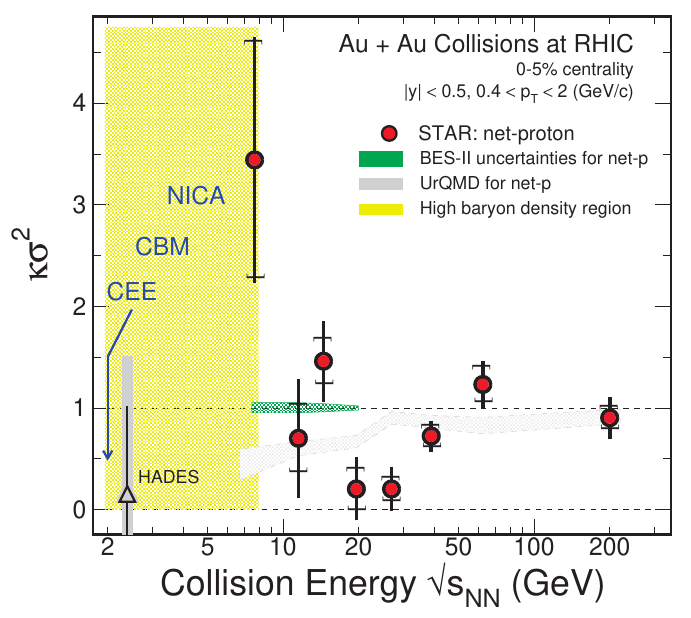} 
\caption{(Color online) Energy dependence of net proton and proton $\kappa\sigma^{2}$ in Au+Au collisions at the center $(0-5\%)$ of RHIC BES-1 ~\cite{Luo2017Search}. The results of the HRG and UrQMD models are shown as dotted lines and gold bands, respectively.}
\label{Net-proton 4th}
\end{figure}

Relativistic heavy ion collisions (HIC) over wide energy range provide the unique way to study QCD phase diagram in terrestrial laboratory.  Over the past two decades, nuclear collisions at the Relativistic Heavy Ion Collider (RHIC) and the Large Hadron Collider (LHC) have collected a large amount of data searching for first-order phase transition signals and precise phase transition CEP.  Between 2010 and 2017, RHIC completed the first phase of the beam energy scanning program (BES-I) ~\cite{Aggarwal2010Higher, Adamczyk2017Bulk}. The second phase (BES-II) was launched in 2018, focusing on the collisions with $\sqrt{S_{NN}} \le 27$ GeV to search sensitive observable  of QCD phase transitions ~\cite{Luo2015Energy, Luo2016Exploring, Abdallah2021Cumulants}, including net proton ~\cite{Stephanov2009NonGaussian, Stephanov2009Sign, Kitazawa2017Properties, Luo2017Search, STAR2021Nonmonotonic}, net charge ~\cite{STAR2014Beam}, net Kaon multiplicity distribution and high order moment analysis ~\cite{LAdamczyk2018Collision}. FIG.\ref{Net-proton 4th} shows the energy dependence of net proton high momentum $\kappa\sigma^{2}$ in Au+Au collisions with the centrality of $0-5\%$  collected in  RHIC BES-I ~\cite{Luo2017Search}. It is observed that  $\kappa\sigma^{2}$ of net proton is closely related to that of proton and exhibits a non-monotonic dependence on collision energy with a significance of $3.1\sigma$ ~\cite{Stephanov2009Sign, STAR2021Nonmonotonic}. In addition, the prediction of UrQMD calculation, which incorporates conservation laws and most of the relevant physics apart from a phase transition and mean field potentials, does not coincide with the trend of the data points at low energies, which means that accurate measurement is very important in the supra-saturation density region and lower beam energies to further reduce experimental errors. In addition to the existing ones, many more experiments are in schedule, like CBM at FAIR, MPD at NICA etc.

In hadron phase, the key component of the QCD phase structure is the equation of state of nucleonic matter (nEOS) ~\cite{BaoAn2014Topical}. For the convenient definition of nEOS,  one writes the nucleon specific energy $E(\rho,\delta)$ of nucleonic matter at zero temperature in quadratic form

\begin{equation}
\label{eos-def}    
E(\rho,\delta)=E_0(\rho)+E_{\rm sym}(\rho)\delta^2
\end{equation}
where $\delta=(\rho_{\rm n}-\rho_{\rm p})/(\rho_{\rm n}+\rho_{\rm p})$ is the isospin asymmetry, $\rho_{\rm n}$ and $\rho_{\rm p}$ are the density of neutrons and protons.  $E_0(\rho)$ is the nucleon specific energy in symmetric nuclear matter and \esym~ is the density dependent nuclear symmetry energy.  Keeping the lowest  order terms, for the symmetric nuclear matter, the nEOS can be written in 

\begin{footnotesize}
\begin{equation}
\label{eos-snm1}    
E_0(\rho)=E_0(\rho_0)+\frac{\kappa_0}{2}\left(\frac{\rho-\rho_0}{3\rho_0}\right)^2+\frac{J_0}{6}\left(\frac{\rho-\rho_0}{3\rho_0}\right)^3+\frac{Z_0}{24}\left(\frac{\rho-\rho_0}{3\rho_0}\right)^4,
\end{equation}
\end{footnotesize}
where the coefficients $\kappa_0$, $J_0$ and $Z_0$ are the incompressibility, skewness and kurtosis parameters, respectively. 

The nuclear symmetry energy is expressed as 
\begin{equation}
\label{eos-snm2}    
 E_{\rm sym}(\rho)=E_{\rm sym}(\rho_0)+L\left(\frac{\rho-\rho_0}{3\rho_0}\right)+\frac{K_{\rm sym}}{2}\left(\frac{\rho-\rho_0}{3\rho_0}\right)^2,
\end{equation}
where $L$ and  $K_{\rm sym}$  are the slope and curvature parameters of the nuclear symmetry energy at normal nuclear matter density $\rho_{0}$, respectively.

In HIC from a few hundred MeV/u to a few GeV/u, nuclear matter above saturation density is created. The density gradients of the compressed nuclear matter will be developed to various forms of flow carrying the information of nEOS, and in return, the flow has been used to probe the incompressibility of nuclear matter ~\cite{Reisdorf2012Systematics}. Recently, HADES collaboration  has published the proton triangular flow $v_{3}$ data at $2.4$ GeV. Compared to UrQMD transport model calculations, it is shown that the proton $v_{3}$ exhibits high sensitivity to the hadron medium EoS ~\cite{Paula2018Directed}. In addition to the flow analysis, the production of $K$ meson,
particularly near threshold, is sensitive to the density of the compressed participant region, and thus is employed as a probe of nEOS.
Thanks to the great efforts from the communities of nuclear physics and astrophysics, the knowledge of $E_0(\rho)$ from $\rho_0$ to a few times of $\rho_0$ is accumulated.

The least known term of nEOS is the density behavior of symmetry energy $E_{\rm sym}(\rho)$. Since the \esym~ gives rise to the isospin diffusion and isospin drift mechanisms, in the terrestrial laboratory, HIC involving neutron rich nuclei provide unique opportunities to constrain \esym~ near and above the saturation density. In a wide energy range,  various isospin probes have been identified and applied to constrain \esym~ near and above $\rho_0$ in the last two decades ~\cite{Bao1997Equation,Xu2000Isospin, Tan2001Fragment,Tsang2001Isotopic,Bao2001Proton,Bao2000Neutron,Xiao2014Probing,Bao2003Isospin}. Some new probes are also introduced recently, including the neutron skin thickness of heavy nuclei ~\cite{Adhikari2021Accurate, Reed2021Implications}, the angular distribution of average $N/Z$ of light charged particles ~\cite{Zhang2017Longtime}, anticorrelation of the ${\rm t/^3He}$ yield ratio with the heavy clusters ~\cite{Yijie2023Observing} for $\rho\approx\rho_0$ and the $\pi^-/\pi^+$ ratio for $\rho>\rho_0$ ~\cite{JEstee2021Probing}. Moreover,  since the observation of the neutron star merging event GW170817, great progress has been made to constrain \esym~ by combining the GW170817 and heavy ion  data ~\cite{Zhou2019Equation, Yangyang2021Insights}. For a recent review, one can refer to  ~\cite{2021BaoProgress}. 

\begin{figure}[!htbp]
\centering
\includegraphics[width=0.48\textwidth]{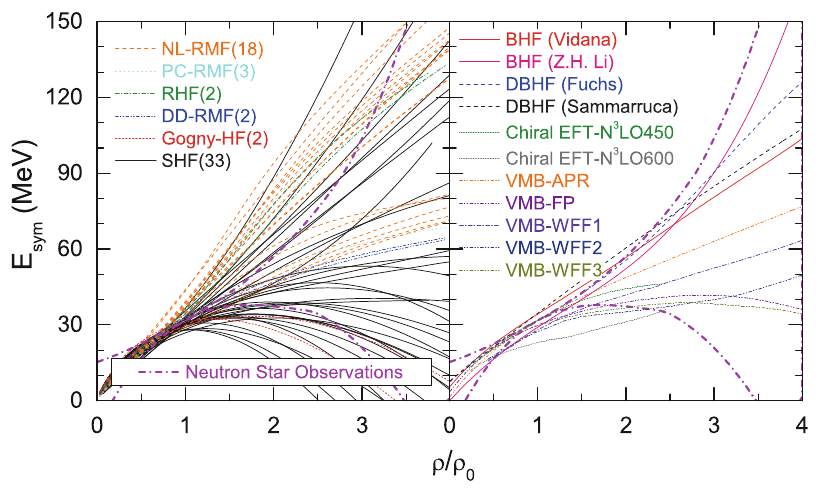}
\caption{(Color online) Research the constraints of symmetry energy in the supra-saturation density region from various theories, and display the constraints on the boundary through neutron star data ~\cite{Bao2019Towards}}
\label{symmetry energy}
\end{figure}

Despite of great progress achieved in the last two decades, still the constraint of \esym~ suffers from large uncertainties ~\cite{Reisdorf2007Systematics, Xiao2009Circumstantial, Xie2013Symmetry, Russotto2011Symmetry, Yangyang2021Insights}, particularly at high density region. World wide efforts in this direction is still ongoing.  In order to finally determine \esym~ in the supra-saturation density region, some new experiments are scheduled or have conducted at some large scientific devices in the world, including but not limited to HADES at GSI, S$\pi$RIT at RIKEN, RAON-LAMPS in Korea etc. These latest experiments will certainly provide better experimental conditions and improve the statistics and reduce the uncertainties. For the recent review, one can refer to the long report ~\cite{Luo2022PropertiesBook}.

To this end, in addition to the RHIC-STAR energy scan experiment, there are several large experimental facilities in operation or construction around the world, which are of great significance for mapping the QCD phase diagrams at high net baryon number density, as well as for constraining the nEOS in hadron phase. In addition to the aforementioned facilities around the world, CEE is currently under construction at the $T_{4}$ Experimental Terminal of the Heavy Ion Research Facility (HIRFL) in Lanzhou ~\cite{LiMing2016Conceptual,Yuan2020Present},  aiming at collecting the first physics data by 2025. The High Intensity Heavy Ion Accelerator Facility (HIAF) is under construction in Huizhou, China ~\cite{SunZhiyu2020Huizhou}. The HIAF in design will be able to provide beams of whole ion species, with the highest  center of mass energy in U+U collisions being  $\sqrt{S_{NN}}=4$ GeV. The upgraded CEE$^{+}$ will be placed in terminal T2 of the HIAF for high energy experiments, providing great opportunities in the studies of QCD phase structure in high baryon density region and nEOS in high density. 

In this paper, the design, performance  and feasibility of various physical goals related to the nEOS studies with CEE are introduced. The simulations  are carried out in the framework of Geant 4. The ultra-relativistic quantum molecular dynamics (UrQMD) model which incorporates the Skyrme potential energy density functional and the in-medium nucleon-nucleon cross section is adopted as the event generator  ~\cite{Li:2011zzp,Wang:2020dru,Yongjia2020Application}. The  analysis and reconstructions of the simulated data are done in CEEROOT platform, which is developed by the collaboration of the CEE experiment, inheriting from Fair ROOT ~\cite{FAIRROOT}. The paper is arranged as following. Section \ref{sec. II} presents the structure of CEE in the technique design and the main functions of each detector subsystem. Section \ref{sec. III} introduce the expected performance of CEE, including the spatial acceptance, momentum resolution, PID ability, and the construction of CEE global trigger signal. Section \ref{sec. IV} presents the verification of  a few  probes sensitive to the nEOS beyond saturation density,  including the yield of light clusters, radial flow, $\pi^{-}/\pi^{+}$ ratio and yield of neutral Kaon mesons.  Section \ref{sec. V} presents a summary.

\section{Structure and simulation of CEE} \label{sec. II}
\subsection{Structure of CEE}\label{sec. II1}

\begin{figure}[htb]
\centering
\includegraphics[width=0.4\textwidth]{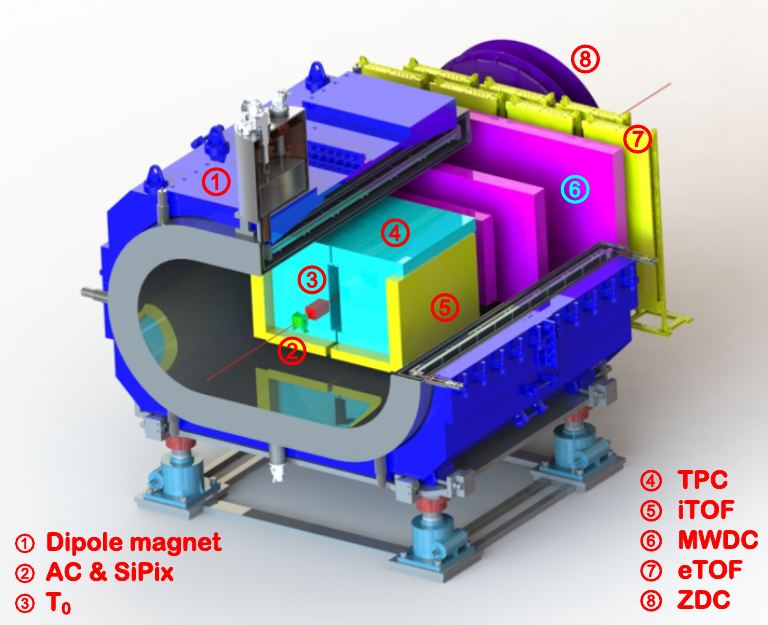}
\caption{(Color online) The design of the CEE}
\label{design of the CEE}
\end{figure}

Aiming at the studies of the nEOS at approximately $2\rho_0$, where $\rho_0$ is the saturation density, CEE is designed to measure the charged particles in nearly $4\pi$ space in center of mass reference ~\cite{LiMing2016Conceptual}.  The design of the CEE is shown in FIG.\ref{design of the CEE}. The main component of CEE is a large-gap dipole magnet, housing the tracking detectors in a nearly homogeneous field of 0.5 T along the vertical direction. The target is located  inside the field, so that the tracking detectors can cover large solid angle. The tracking detectors  include two time projection chambers (TPC) ~\cite{Li2016Simulation} surrounded by the inner time-of-flight (iTOF) detectors covering the midrapidity ~\cite{Wang2022CEE}, and an array consisting of three multiwire drift chambers (MWDC) followed by an end cap time-of-flight (eTOF) wall for the measurement of charged particles at forward rapidity  ~\cite{Botan2020eTOF}. At the downstream  end of the CEE spectrometer, there is an array of plastic scintillators to measure the charged particles near zero degree (ZDC) ~\cite{Zhu2021Prototype}. The start timing detector ($\rm T_0$) ~\cite{Hu2017T0, Hu2020Cooling} and the active collimator (AC) are placed on the beam line in the upstream side of the target. Besides, there is also a silicon pixel positioning detector (SiPix) ~\cite{Hulin2022Design, Liu2023Design} to monitor the beam position at the upstream side to $\rm T_0$.  Tab.\ref{technical indicators of CEE} shows the main technical performance of CEE system.

\begin{table}[!htbp]
\caption{(Color online) Technical indicators of CEE.}
\label{technical indicators of CEE}
\setlength\tabcolsep{13.5pt}
\centering
\footnotesize
\begin{tabular}{cc}
\toprule
Item & value \\[1.0pt] 
\midrule
Maximum beam energy & $\rm {0.5 GeV/u (U) - 2.8 GeV (p)}$ \\[1.0pt]
Bean type & $p \sim U$ \\[1.0pt]
Maximum event rate & $\rm 10~kHz$ \\[1.0pt]
Acceptance & $>50\%$ \\[1.0pt]
Total channel number & $\rm 20 k$ \\[1.0pt]
\bottomrule
\end{tabular}
\centering
\footnotesize
\end{table}

The tracking detector at midrapidity are the two identical TPCs surrounded by iTOF walls. The two TPCs are installed in a left-right symmetric  mode in order to leave a 20 cm space in between for the beam passing through.  The target position is in the gap of the two TPCs, being 15 cm to the rear surface of the field cage. The electronics of the TPCs are placed on the top plane above the two TPCs. The high voltage is fed into the field cage from the bottom plane. A high voltage step-down structure surrounded by teflon blocks is designed to avoid a sharp voltage decrease near the HV point.  The field cages are made of Kapton layer with copper electrodes printed on both surfaces, ensuring the distortion of the electric field to be less than $0.1\%$. A three-layer large-area gaseous electron multiplier (GEM) foils are stacked on the top of the sensitive volume of TPC to amplify the arrived electrons and then induce signals on the readout pads. The  signals of each pad  are readout by the specially designed front end electronics (FEE)  and then transferred to the digitization board adopting SAMPA chips ~\cite{Adolfsson2017SAMPA, Jiangyue2023Development}. 

The forward tracking detectors consist of three MWDCs and an eTOF wall placed at forward angle. Two sets of MWDC are placed inside the field of the dipole and the third one followed by the eTOF wall are  placed outside the magnet. On each MWDC and the eTOF wall, an inactive area is designed for the beam particles to pass through without inducing enormous charges in the sensitive volume of the detectors. Each MWDC contains six layers of sensitive wires in three directions, called X, U  and V wires, meeting $0^\circ$, $30^\circ$ and $-30^\circ$ with respect to the vertical direction, respectively. For each direction of wire, there are two layers of drift cells, displacing by half size of a cell in order to discriminate the ghost hit. Both the entrance and exit windows of the MWDCs are made of aluminum coated Mylar foil. The detector operates in atmospheric flow-gas mode. The high voltage are fed to the anode wires directly. The signals, readout with a capacitance coupling, is amplified and shaped in the FEE before being transferred to a SCA chip  for digitization. Then the digitized data are sent to the BDM module, where the timing and amplitude information is online computed and transferred to data acquisition system (DAQ) system for saving. Table.\ref{technical indicators of TPC and MWDC} lists the main geometric dimensions and technical specifications of the TPC and MWDC.

\begin{table*}[!htbp] 
\caption{(Color online) The geometry size and technical indicators of TPC and MWDC}
\label{technical indicators of TPC and MWDC}
\setlength\tabcolsep{10pt}
\centering
\footnotesize
\begin{tabular}{cc|cc}
\toprule 
\multicolumn{2}{c}{TPC parameters} & \multicolumn{2}{c}{MWDC parameters} \\[1.0pt] 
\midrule
Item & value & Item & value \\[1.0pt] 
\midrule
Effective sensitive area (half) & 500(x) $\times$ 800(y) $\times$ 900(z) mm$^{3}$ & MWDC1 effective sensitive area & 136(x) $\times$ 63(y) cm$^{2}$ \\[1.0pt] 
Channel number & $\geq$ 10000 & MWDC2 effective sensitive area & 190(x) $\times$ 88(y) cm$^{2}$ \\[1.0pt] 
Sensitive area of GEM film & 900 mm $\times$ 500 mm & MWDC3 effective sensitive area & 274(x) $\times$ 126(y) cm$^{2}$ \\[1.0pt] 
Track multiplicity & $<$ 100 & Channel number & $\geq$ 3000 \\[1.0pt] 
Maximum event rate & 1kHz & Transverse position resolution & 0.3 mm \\[1.0pt] 
Maximum field strength & 400 V/cm & Track reconstruction efficiency & $\sim98\%$ \\[1.0pt] 
Momentum resolution & $\sim5\%$ & Momentum resolution & $\sim5\%$ \\[1.0pt] 
\bottomrule 
\end{tabular}
\end{table*}

The time-of-flight (TOF) signals are measured by  T$_{0}$, inner TOF wall (iTOF) and the outer TOF wall (eTOF), respectively. The T$_{0}$ placed in front of the target is made of  scintillator foil read out by SiPM from the side, delivering the initial time of heavy ion collision. Both iTOF and eTOF, consisting of multi-gap resistive plate chambers (MRPC),  are installed to record the arrival time of the  charged particles passing through TPC and MWDC, respectively. Thus, combining the tracking  and TOF detectors, one can deduce the PID information  via $dE/dx - p/Q$ in the tracking detectors and $p/Q-\beta$, respectively, where p is the momentum and $\beta$ is the velocity  calculated from the track length and TOF. Table.\ref{The technical indicators of iTOF and eTOF} shows the main performances of TOF system. The total area of the iTOF and eTOF systems is about 12 m$^{2}$. In iTOF and eTOF, the hitting position can also be derived  through the time difference of the signals at both ends. 

The iTOF, grouped by modules,  surrounds the left, right and bottom surfaces of the TPC. The upper surface is not covered by TOF detectors because  the read out electronics of the TPC is arranged on the top.  The left and right iTOF walls are composed of six iTOF modules, each iTOF module contains three MRPC detectors, and the left and right envelope size is 100(x) $\times$ 1000(y) $\times$ 1355(z) mm$^3$. The bottom iTOF wall contains two symmetrical parts, consisting of 6 iTOF modules and 12 MRPC detectors, with the bottom envelope size of 670(x) $\times$ 1355(z) $\times$ 100(y) mm$^3$. The eTOF wall is located at the down stream end of the dipole magnet, being 2.6 m to the target position. Each MRPC has  10 $\times$ 0.25mm air gaps and self-sealing technology is applied for the first time. The whole wall is divided into 7 modules, each module contains $3 \sim 4$ self-sealed MRPC ~\cite{Botan2020eTOF}.  The signals of the TOF detectors are collected and amplified  by  the MRPC special electronics chip NINO ~\cite{Anghinolfi2003NINO} in FEE, and then the high precision time information is obtained through the high performance time digitization module (TDM) ~\cite{Lu2021Readout}.

\begin{table*}[!htbp] 
\caption{(Color online) The technical indicators of iTOF and eTOF}
\label{The technical indicators of iTOF and eTOF}
\setlength\tabcolsep{13.5pt}
\centering
\footnotesize
\begin{tabular}{cc|cc}
\toprule 
\multicolumn{2}{c}{iTOF parameters} & \multicolumn{2}{c}{eTOF parameters} \\[1.0pt] 
\midrule
Item & value & Item & value \\[1.0pt] 
\midrule
Total effective area & 6 m$^{2}$ & Total effective area & 3.2 $\times$ 1.6 m$^{2}$ \\[1.0pt] 
Channel number & $\sim$ 2400 & Channel number & $\sim$ 1500 \\[1.0pt] 
Average count rate of single channel & 10kHz & Average count rate of single channel & 10kHz \\[1.0pt] 
Time resolution accuracy & $<$ 50ps & Time resolution accuracy & $<$ 60ps \\[1.0pt] 
\bottomrule 
\end{tabular}
\end{table*}

\begin{figure}[!htbp]
	\centering 
	\subfigure{
        \includegraphics[width=0.50\linewidth]{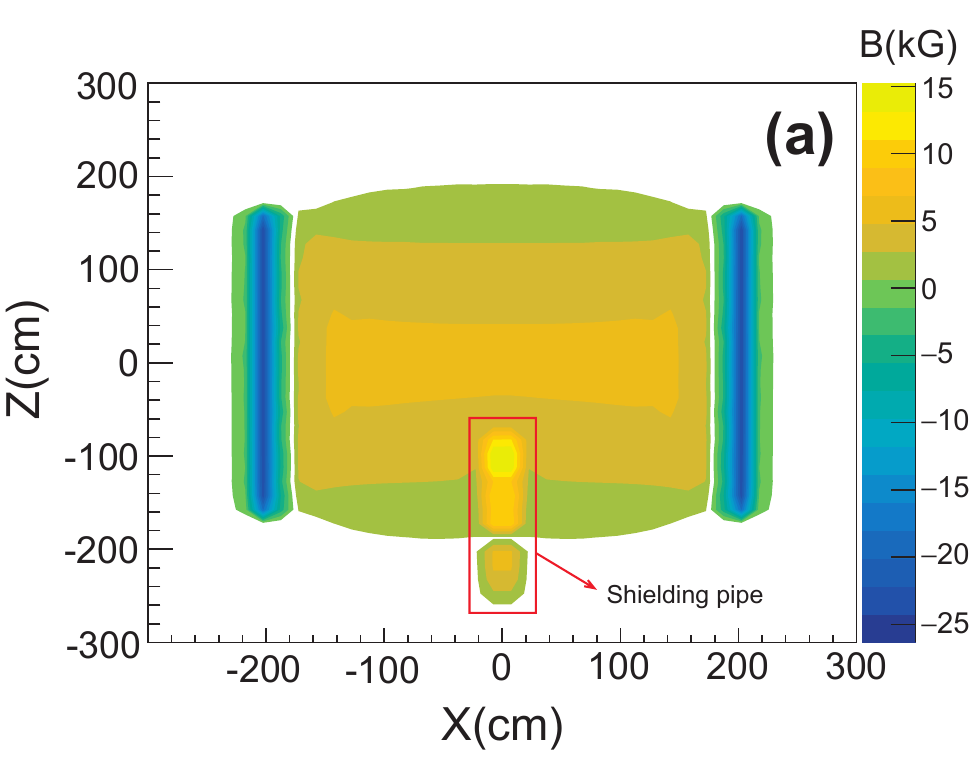}}
	\subfigure{
		\includegraphics[width=0.45\linewidth]{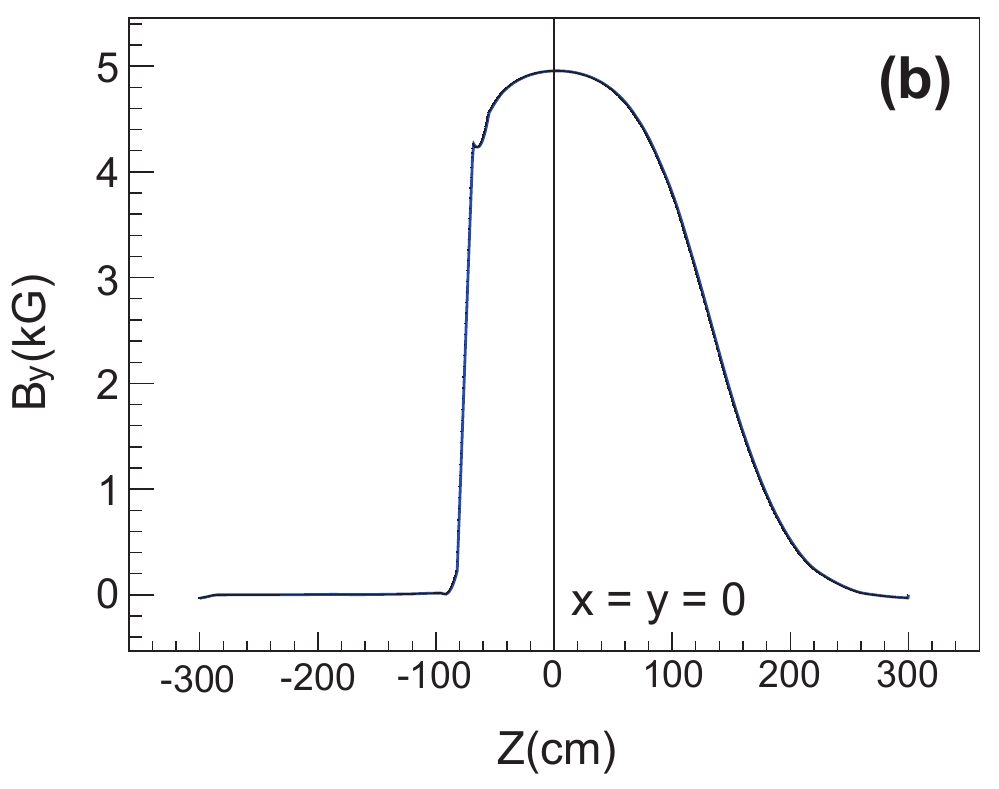}}
	\caption{(Color online) The magnetic field distribution in the center of the secondary magnet. The left (a) shows the magnetic field distribution in the X-Z plane, and the magnetic field shield tube is shown in the red box; The right figure (b) shows the distribution of magnetic field intensity in the y direction in the beam direction}
	\label{magnetic field} 
\end{figure}

The active collimator functioning to veto the off-target reaction background  is placed between $T_0$ and the target. It is composed by  four plastic scintillators ( 250 $\times$ 250 mm$^{2}$) distributed around the entrance of the target chamber, leaving a 50 $\times$ 50 mm$^{2}$  hole in the center. The SiPM is used to read out the signals. 

The ZDC is mainly used to measure the spectators from HIC, and to determine the centrality and the reaction plane of collisions. The ZDC is located at the most downstream position of the whole spectrometer, about 300 cm to the target.  The overall transverse layout of ZDC is rotationally symmetric to the beam line. It is divided into 24 equal sectors, each of which contains 8 trapezoid scintillators placed in row along the radial direction. The outer radius of the ZDC sensitive area is 100 cm while the inner radius is about 10 cm, through which the non-reacted beam particles can pass. 

The large gap superconducting dipole magnet is to bend the charged particles produced in the projectile-target collisions. The raw dimension of the magnet is 2700(L) $\times$ 4300 (W) $\times$ 2.7 (H) mm$^{3}$. The space of the uniform field housing the sensitive volume of the TPC  is 900 (L) $\times$ 1200 (W) $\sim$ 800 (H) mm$^{3}$. The central field  is 0.5 T in strength with a uniformity  better than $5\%$. FIG.\ref{magnetic field} shows the magnetic field distribution in the center of the dipole magnet.

In addition, the CEE also includes the following supporting systems. The DAQ uses D-Matrix architecture to provide data storage. The main function of the clock system is to provide high precision global synchronous clock for the electronics modules and  DAQ  system. The trigger system provides global trigger  signals in beam experiments or in detector test run. Refer to Section \ref{sec. III3} for details about the construction of the global trigger signals of CEE.

\subsection{Event generator and simulation tool}\label{sec. II2}

In order to constrain nEOS, particularly for supra-saturation density, one usually use the given  $E_0(\rho)$  and \esym~ as input  in transport model for HIC or in solving the Tolman–Oppenheimer–Volkoff (TOV) equations for neutron star structures. The predicted results are then compared to experimental data or astrophysical observations. By adjusting the parameters of  $E_0(\rho)$  and \esym~ until the results converges to the experimental results, one arrives a constraint of nEOS. And in return, transport model is usually employed as the event generator in the feasibility studies of nEOS via in terrestrial heavy ion experiment.   

UrQMD model is a transport model widely used to describe the relativistic heavy ion collisions and proton induced reaction process ~\cite{Bass1998Microscopic, MBleicher1999Relativistic}, ranging from sub-GeV/u to TeV/u energies covered by SIS, AGS, SPS, RHIC and LHC. In order to apply the UrQMD model as an event generator for CEE, the model is adopted by improving the isospin-dependent nuclear potential and collision terms.

The UrQMD transport model in this work is provided by Huzhou University in China ~\cite{Li:2011zzp,Wang:2020dru,Yongjia2020Application}, and the main improvements include: 1) The mean-field potential
derived from the Skyrme potential energy density functional has been implemented into ~\cite{Wang:2013wca,Wang:2014rva,Yongjia2020Application};
2) The nuclear medium effects on $NN$ cross sections are considered, the density-, momentum- and isospin-dependent $NN$ cross section is adopted~\cite{Li:2006ez,Li:2011zzp,Li:2018wpv,Li:2022wvu};
3) The isospin-dependent minimum spanning tree (iso-MST) method \cite{Zhang:2012qm} is adopted to construct clusters. 
In addition, the yield and flow observables about $\pi$ and $K$, which are effective and sensitive probes of symmetry energy, can be better reproduced by introducing proper pion-nucleon potential and kaon-nucleon potential \cite{Du:2018ruo,Liu:2018xvd,Yangyang2021Insights}. 
The effective two-body interaction potential energy $U$ is decomposed into the Coulomb energy $U_{\rm{Coul}}$, the Skyrme potential energy $U_{\rho}$ and the momentum-dependent potential energy $U_{\rm{md}}$. The $U_{\rho}$ and $U_{\rm{md}}$ is written as $U_{\rho,\rm{md}}=\int u_{\rho,\rm{md}}\rm{d}\bf{r}$, where 

\begin{equation}\label{eqn-1}
   \begin{aligned}
    u_{\rho}=
	&\frac{\alpha}{2}\frac{\rho^2}{\rho_{0}}+\frac{\beta}{\eta+1}\frac{\rho^{\eta+1}}{\rho_{0}^{\eta}}+\frac{g_{\text{sur}}}{2\rho_{0}}(\nabla\rho)^{2} \\
	&+\frac{g_{\text{sur,iso}}}{2\rho_{0}}[ \nabla(\rho_{n}-\rho_{p})] ^{2}+u_{\rho}^{\rm{sym}},
    \end{aligned}
\end{equation}
and
\begin{equation}
u_{\rm{md}}=t_{\rm{md}} \ln^{2}\left[1+a_{\rm{md}} (\textbf{p}_{i}-\textbf{p}_{j})^2\right] \frac{\rho}{\rho_{0}}.
\end{equation}

The symmetry potential energy density functional reads as
\begin{equation}
u_{\rho}^{\rm{sym}}=\left [ A_{\text{sym}}(\frac{\rho}{\rho_{0}})+B_{\text{sym}}(\frac{\rho}{\rho_{0}})^{\gamma}+C_{\text{sym}}(\frac{\rho}{\rho_{0}})^{5/3}\right ]\delta^2\rho.
\end{equation}
For the Skyrme interactions, the symmetry potential energy terms come from two-body, three-body interaction terms, and its related parameters are $A_{\text{sym}}$, $B_{\text{sym}}$, and $C_{\text{sym}}$, respectively. All the parameters
used in the UrQMD model, such as $\alpha$, $\beta$, $\eta$, $g_{\rm{sur}}$, $g_{\rm{sur,iso}}$, $A_{\text{sym}}$, $B_{\text{sym}}$, $C_{\text{sym}}$ can be derived from the standard Skyrme parameters, $x_{0}$, $x_{1}$, $x_{2}$, $x_{3}$, $t_{0}$, $t_{1}$, $t_{2}$, $t_{3}$, $\sigma$ \cite{Wang:2013wca,Wang:2014rva,Liu:2020jbg}. In addition, one can set $A_{\text{sym}}=C_{\text{sym}}=0$ and $B_{\text{sym}}=C_{s}/2$, and adopt the density power
law form to investigate the density-dependent symmetry energy\cite{YingXun2020Progress,Liu:2020jbg}
\begin{equation}
u_{\rho}^{\rm{sym}}=\left[ \frac{C_{s}}{2}(\frac{\rho}{\rho_{0}})^{\gamma} \right]\delta^2\rho.
\end{equation}

\begin{figure}[htb]
\centering
\includegraphics[width=0.45\textwidth]{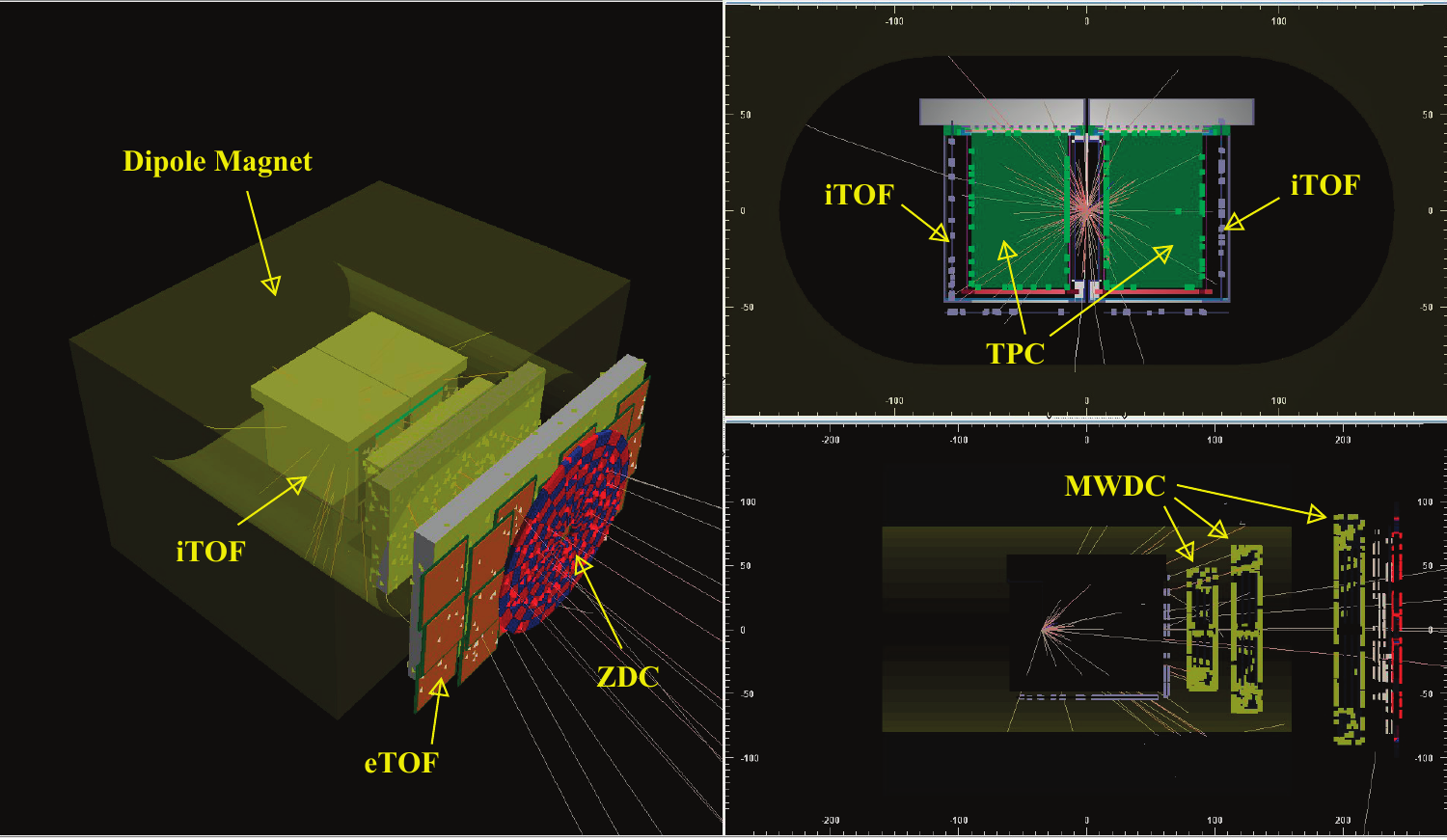}
\caption{(Color online) The detector structure and the  event display of CEE in fast simulation.}
\label{CEE Display}
\end{figure}

The simulations and the reconstructions of CEE experiment is implemented in the framework called CEEROOT, a platform inheriting from Fair ROOT software ~\cite{FAIRROOT} based on the ROOT and Geant 4 packages. User routines are developed to complete the data calibration, event reconstruction, efficiency correction and histograming. In CEEROOT, the  subdetectors, the dipoles and the main materials are constructed using Geant 4 class libraries. When the particles produced in HIC provided  by the event generator are input to the simulator, the propagation of the particles and the responses of the detectors are simulated. Then the signals in the sensitive volume of the detectors are digitized and saved to hard disk for further reconstruction. In this studies, the detector systems, including the magnetic field, T$_{0}$, iTOF, eTOF, TPC, MWDC and ZDC are constructed as designed. The improved UrQMD model described in Section.\ref{sec. II1} is used as the event generator.  FIG.\ref{CEE Display} shows the event display and detector configuration in the simulations. The detector performance based on fast simulations  will be introduced in the next section.

\section{CEE performance study} \label{sec. III}
This section presents the fast simulation studies on CEE performance. Using a symmetric collision $^{208}$Pb+$^{208}$Pb, the acceptance, the PID ability and the centrality selectivity are discussed. The design of the global trigger design in heavy ion collision  of CEE is introduced. 

\subsection{Acceptance}\label{sec. III1}

\begin{figure}[htb]
\centering
\hspace{-0.5cm}\includegraphics[width=0.46\textwidth]{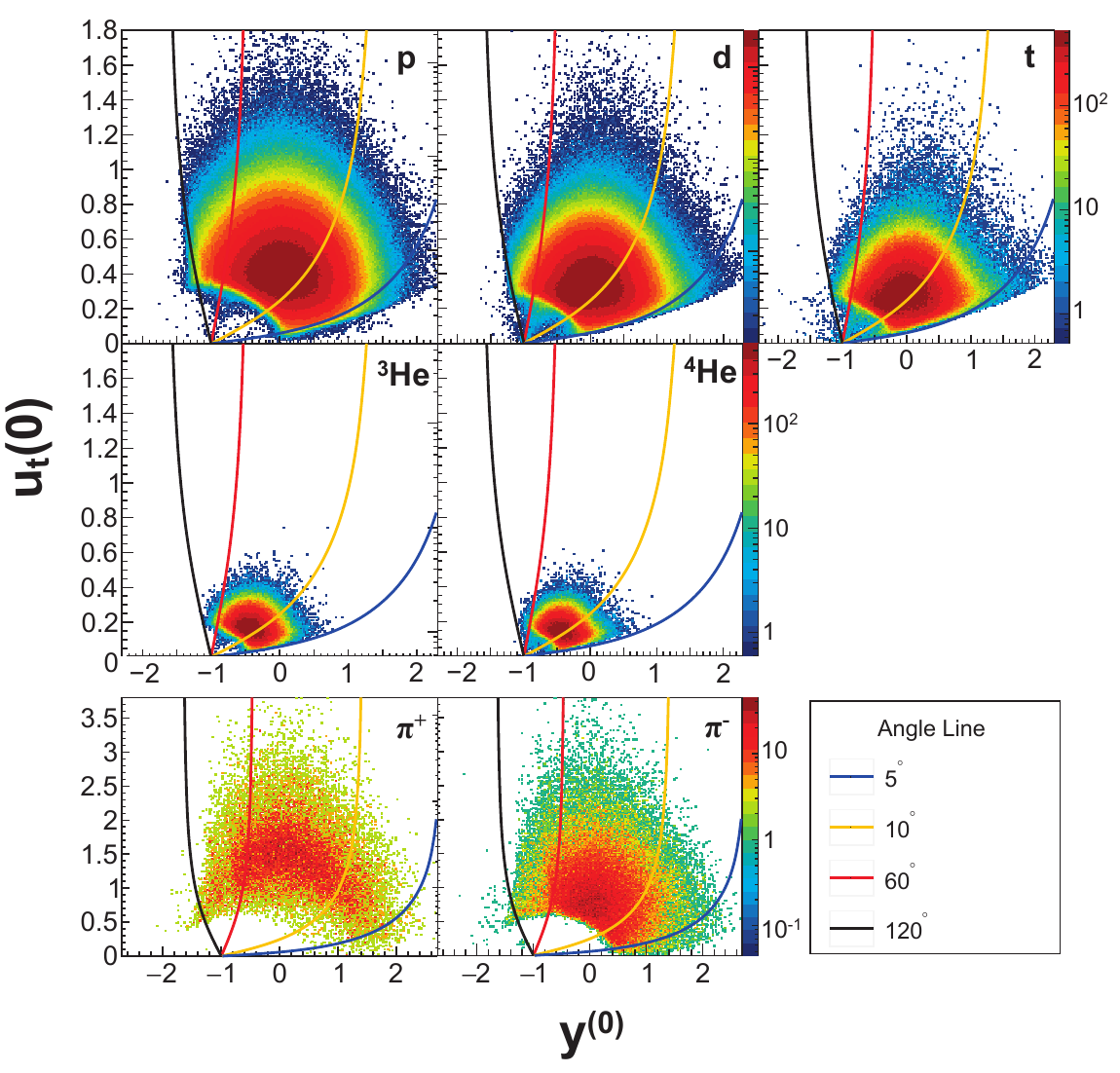}
\caption{(Color online) The phase space distribution of various produced particles in the $^{208}$Pb+$^{208}$Pb collision at 0.3 GeV/u.}
\label{PbPb phase space}
\end{figure}

It is usually required to measure nearly the whole products in the studies of nEOS and QCD phase structure, naturally one favors large acceptance for CEE.  FIG.\ref{PbPb phase space} shows the phase space distribution of different kinds of particles at $b=0$ fm in $^{208}$Pb+$^{208}$Pb collision at 0.3 GeV/u. The abscissa is the reduced rapidity defined by $y^{(0)}=y_{\rm lab}/y_{\rm cm}-1$, where $y_{\rm lab}$ is the rapidity of particles in the laboratory reference, and $y_{\rm cm}$ is the rapidity of the center of mass of the collision system. The ordinate corresponds to the reduced transverse velocity  $u_{\rm t}^{(0)}=p_{\rm t}/(m\beta_{\rm b}\gamma_{\rm b})$, where $p_{\rm t}$ and $m$ represent the transverse momentum and the mass of the particles, respectively. Here $\beta_{\rm b}$ and $\gamma_{\rm b}$ are the beam velocity and the corresponding Lorentz factor in laboratory. The curves corresponding to  different angles in laboratory reference are plotted on top of the data. The MWDC array covers the range of $\theta_{\rm lab}<30^\circ$, while the TPC mainly covers $\theta_{\rm lab}>30^\circ$. It is shown that more than 90\% of the phase space in center of mass can be covered, particularly for the symmetric systems because one can reflect the distribution with respect to  $y^{(0)}=0$ following the physical requirement. 

\begin{figure}[htb]
\centering
\hspace{-0.5cm}\includegraphics[width=0.48\textwidth]{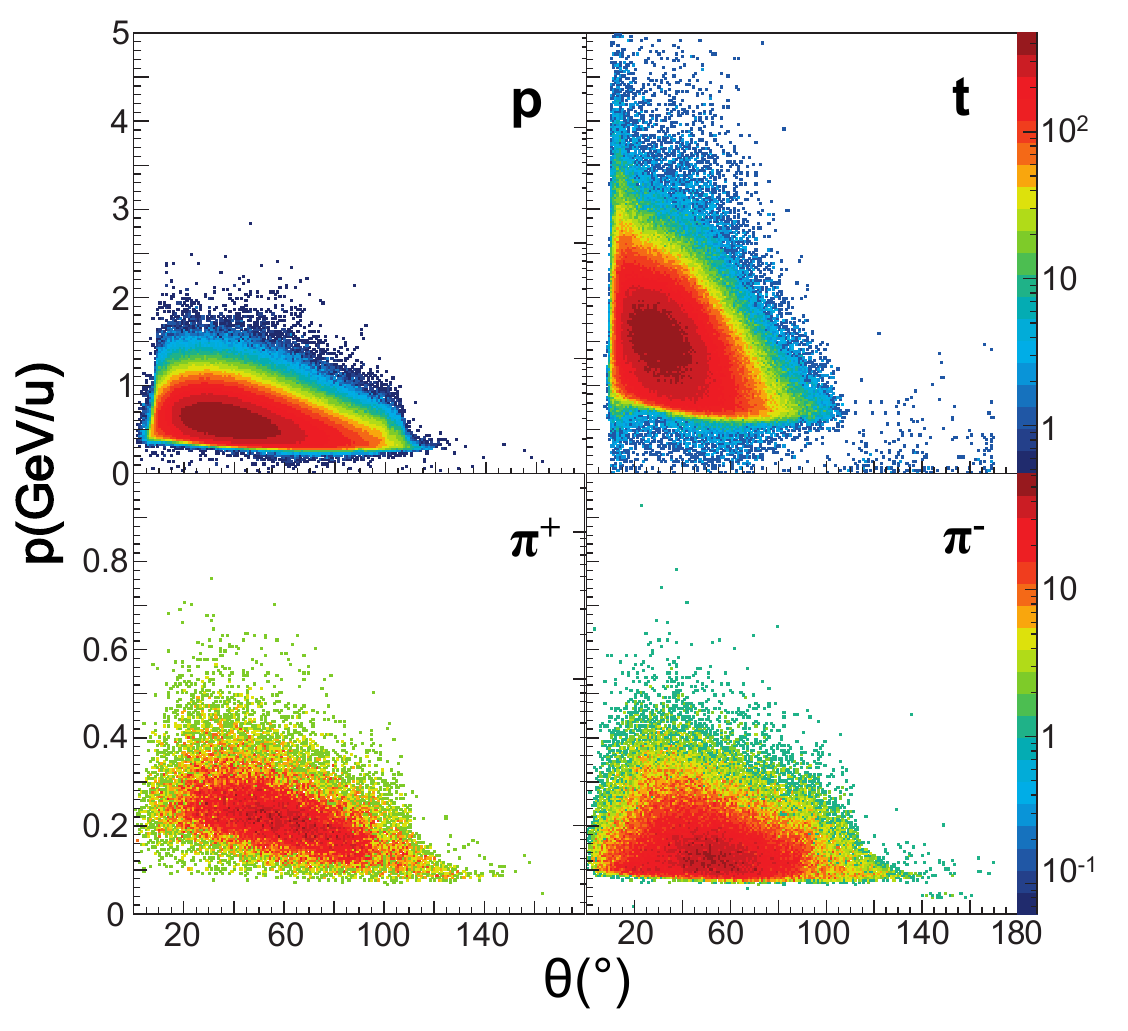}
\caption{(Color online) Correlation between the momentum $p$ and polar angle $\theta_{\rm lab}$ in laboratory for  protons, tritium, $\pi^{+}$ and $\pi^{-}$ in the reactions of $^{208}$Pb+$^{208}$Pb at 0.3 GeV/u.}
\label{p-theta}
\end{figure}

FIG.\ref{p-theta} shows the scattering plot of the momenta $p$ {\it vs.} $\theta_{\rm lab}$ for protons, tritium, $\pi^{+}$ and $\pi^{-}$ in the reaction of ${}^{208}$Pb+${}^{208}$Pb at 0.3 GeV/u in  laboratory reference system. The distribution only pass the filter of TPC and MWDC with a track length cut $L_{trk} >30$ cm, while no conditions of the TOF hits are applied. Again, the wide angular acceptance of the tracking detectors are illustrated.

\begin{figure}[htb]
\centering
\hspace{-0.5cm}\includegraphics[width=0.46\textwidth]{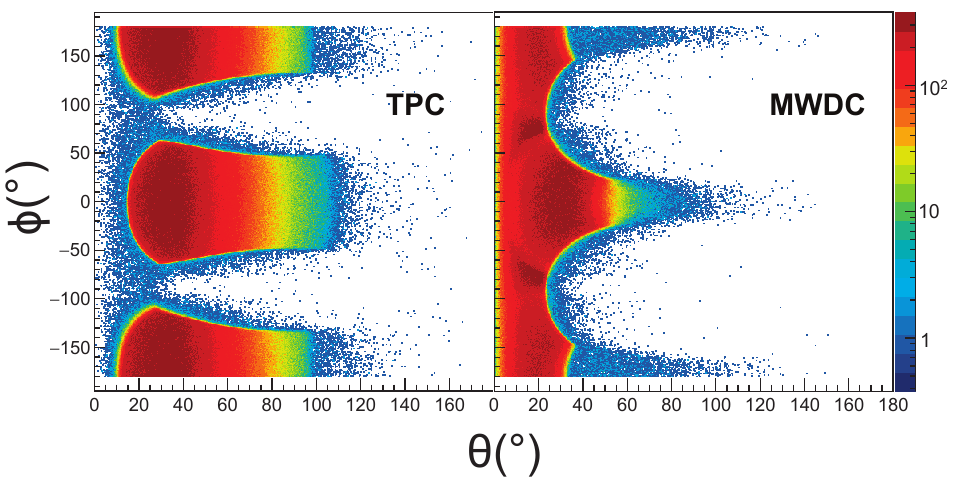}
\caption{(Color online) Angular coverage of proton by the tracking detectors TPC (a) and MWDC (b). }
\label{phi-theta}
\end{figure}

FIG.\ref{phi-theta} further presents the angular coverage of the tracking detector TPC and MWDC on $\theta-\phi$ plane for protons as an example. Particles with a track length shorter than 30 cm in the TPC have been cut. While the polar angle is well covered at $\theta_{\rm lab}<100^\circ$, there is obvious efficiency lose in azimuth. Namely in the vicinity of $\phi\approx 90^\circ$ and  $\phi\approx 270^\circ$, both detectors lose the detection ability.

\subsection{Particle identification}\label{sec. III2}

\begin{figure}[htb]
\centering
\includegraphics[width=0.50\textwidth]{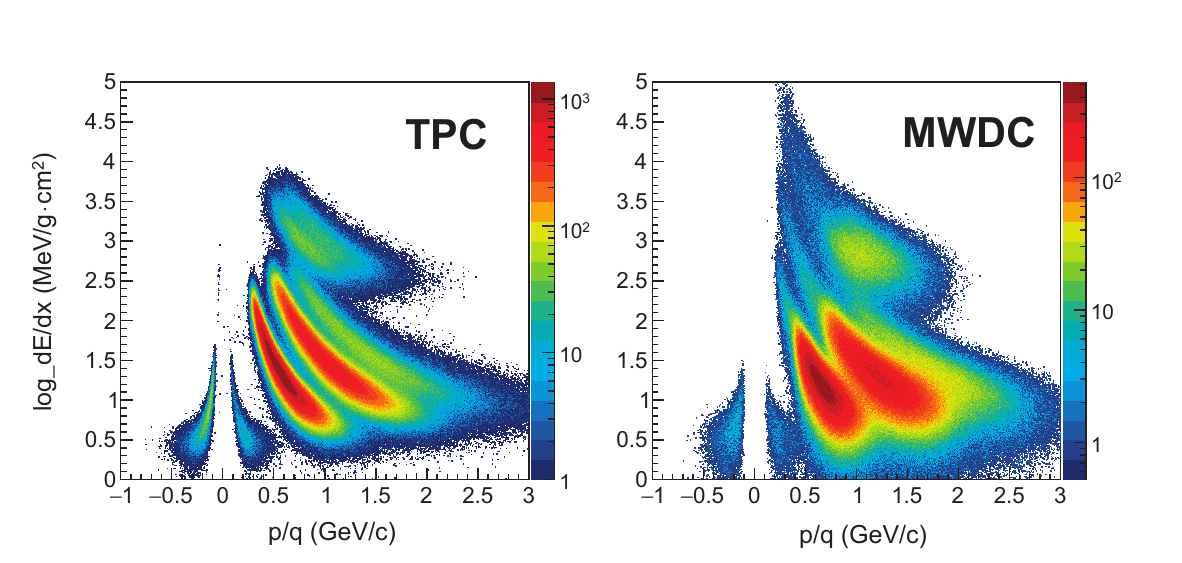}
\caption{(Color online) The predictions of particle identification in TPC and MWDC using the energy loss $dE/dx$ as a function of magnetic rigidity $p/q$.}
\label{PID}
\end{figure}

Since CEE has two tracking detectors, TPC and MWDC, and TOF detectors, both methods using $dE/dx-p/Q$ and $p-TOF$ can be applied, respectively. In the simulation studies, for the  $dE/dx-p/Q$ correlation obtained by TPC and MWDC,  the total energy loss $dE$ is accumulated in each step when a particle is propagating in the sensitive volume of TPC or MWDC filled with the working gas of ($\rm {90\% ~Ar + 10\%~ CO_{2}}$). When a complete cell is processed, the energy loss rate of $dE/dx$ is calculated by summing the energy losses in all steps and smeared by a given variance. Since the distribution of  $dE/dx$ follows Landau formula showing divergence at large value, usually a high energy cut is applied to fill the histogram. The momentum $p$ is obtained by fitting the hits generated by the ionization of the energetic particles in  propagation. Tracking finding is not yet implemented, so the fitting is done directly to each individual track and a smearing is introduced directly  according to the position resolution of the tracking detector. Fig. \ref{PID} presents the $dE/dx-p/Q$ for TPC and MWDC, respectively. It is shown that different isotopes of $Z=1$ and $Z=2$ can be well identified in TPC. While in MWDC, due to the total sampling times is only 18, corresponding to 18 layers of anode wire planes, the determination of the momentum is moderately degraded, although the tracking residue is controlled in $300~\mu$m.  

\begin{figure}[htb]
\centering
\hspace{-0.5cm}\includegraphics[width=0.45\textwidth]{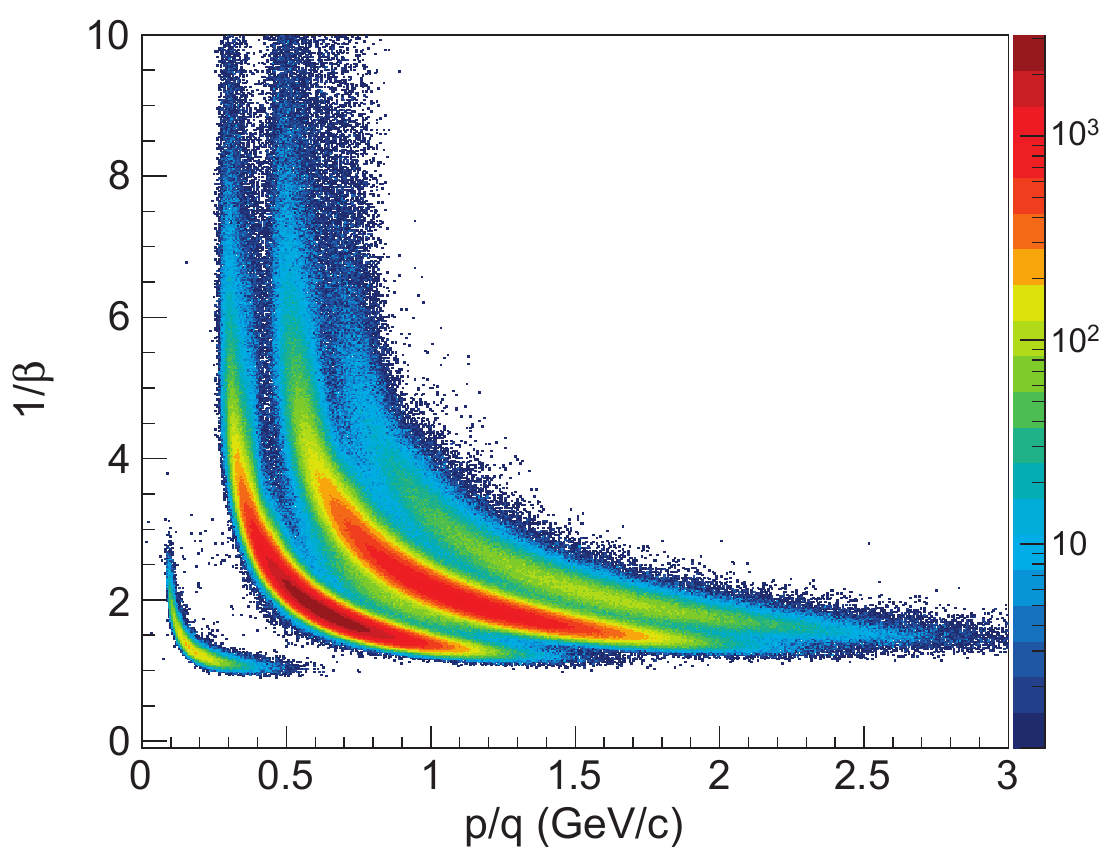}
\caption{(Color online) The change of charged particle flight velocity with momentum measured by iTOF}
\label{beta iTOF_TPC}
\end{figure}

With increasing of $p/q$, the separation between neighbouring bands on the $dE/dx-p/q$ is reduced, particularly between pions and protons beyond 1 GeV/c. Thus, $p-{\rm TOF}$ correlation is applied to improve the PID ability. Once the track length $L_{\rm trk}$ is determined by the tracking detectors, the speed of the particle can be calculated and the mass $m$ can be resolved at a given momentum $p$ by ~\cite{Klempt1999Review} 
\begin{equation}\label{eqn-3} 
    m=\frac{p}{c}\sqrt{\frac{c^{2}t^{2}}{L^{2}_{\rm trk}}-1}
\end{equation}
where $c$ is the speed of light. This method can be applied for both configurations of TPC+iTOF and   MWDC+eTOF, respectively. To achieve the intended performance, the timing resolutions of iTOF and eTOF are designed as 50 ps and 60 ps, respectively. FIG.\ref{beta iTOF_TPC} presents the velocity of the particles as a function of the momentum $p$ measured in TPC. It is shown that the pions and $Z=1$ isotopes are well identified.

\begin{figure}[!htbp]
	\centering 
	\subfigure{
        \hspace{-0.8cm}\includegraphics[width=0.52\linewidth]{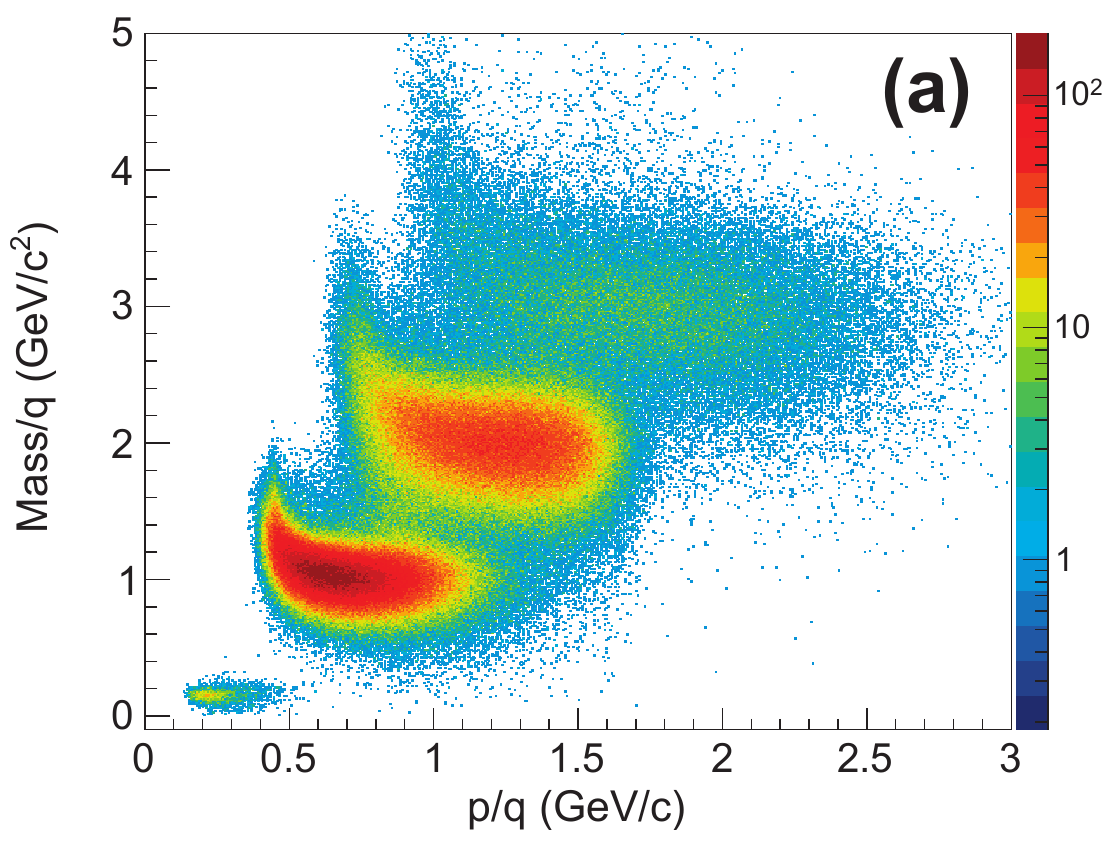}}
	\subfigure{
		\includegraphics[width=0.45\linewidth]{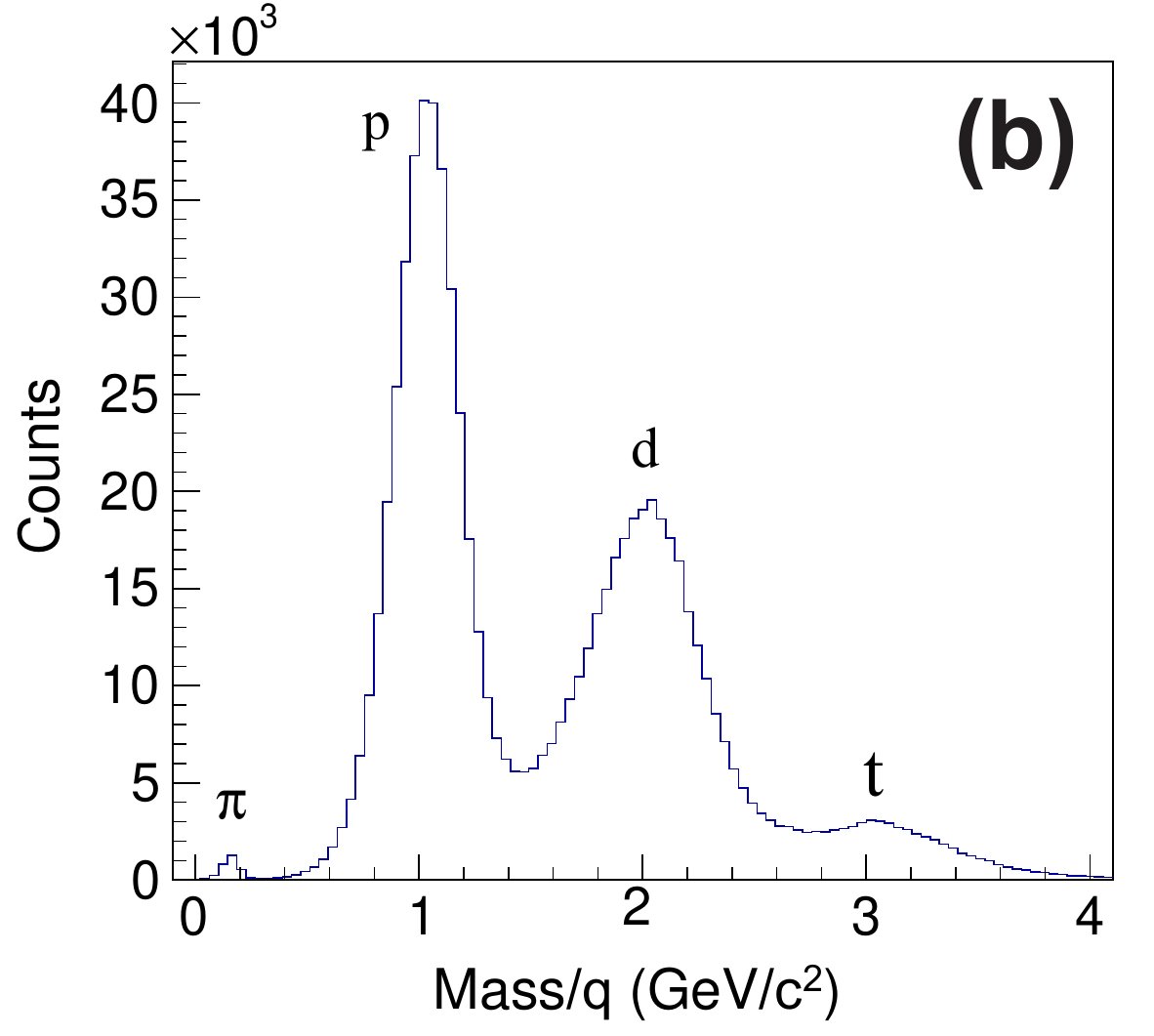}}
	\caption{(Color online) (a) Mass $m$ and momentum $p$ scattering plot obtained by eTOF  and MWDC, (b) mass distribution spectrum.}
	\label{eTOF MWDC Mass} 
\end{figure}

FIG.\ref{eTOF MWDC Mass}(a) shows the mass and momentum correlation by combining the eTOF delivering the time of flight and MWDC providing the momentum information.  The mass distribution is further plotted in  FIG.\ref{eTOF MWDC Mass}(b).  It can be seen that at the beam energies of interest, the light charged particles with $Z=1$ can be separated clearly, although a small distortion is occurred at the low momentum border due to the influence of the gap of the detectors, which is not corrected in the track length estimation.

The resolution of the momentum reconstruction of the tracks in $\vec B $ field is determined by the hit position resolution and the track length $L$ in the bending area, written as  

\begin{footnotesize}
\begin{equation}\label{eqn-2} 
    (\frac{\sigma_{p}}{p})^{2}=(\frac{720}{N+4}\frac{\sigma_{\rm xz}p_{\rm xz}}{0.3BL^{2}_{\rm trk}})+(\frac{0.0523}{\beta B\sqrt{L_{\rm trk}X_{0}}\sin\theta})+(\cot\theta\sigma_{\theta})^{2}
\end{equation}
\end{footnotesize}

 where $B$ is the magnetic field strength and  $\theta$ is the polar angle. N is the number of samplings, $\sigma_{\rm xz}$ and $\sigma_{\theta}$ are the hit position resolution in x-z plane perpendicular to the field  $\vec B$ and the angular resolution, respectively. $X_{0}=1.2$ m is the radiation length.
 Given the hit position uncertainty of the tracks in the tracking detectors, the resolution of the momentum reconstruction can be investigated by fast simulations.  
 
\begin{figure}[htb]
\centering
\hspace{-1.0cm}\includegraphics[width=0.5\textwidth]{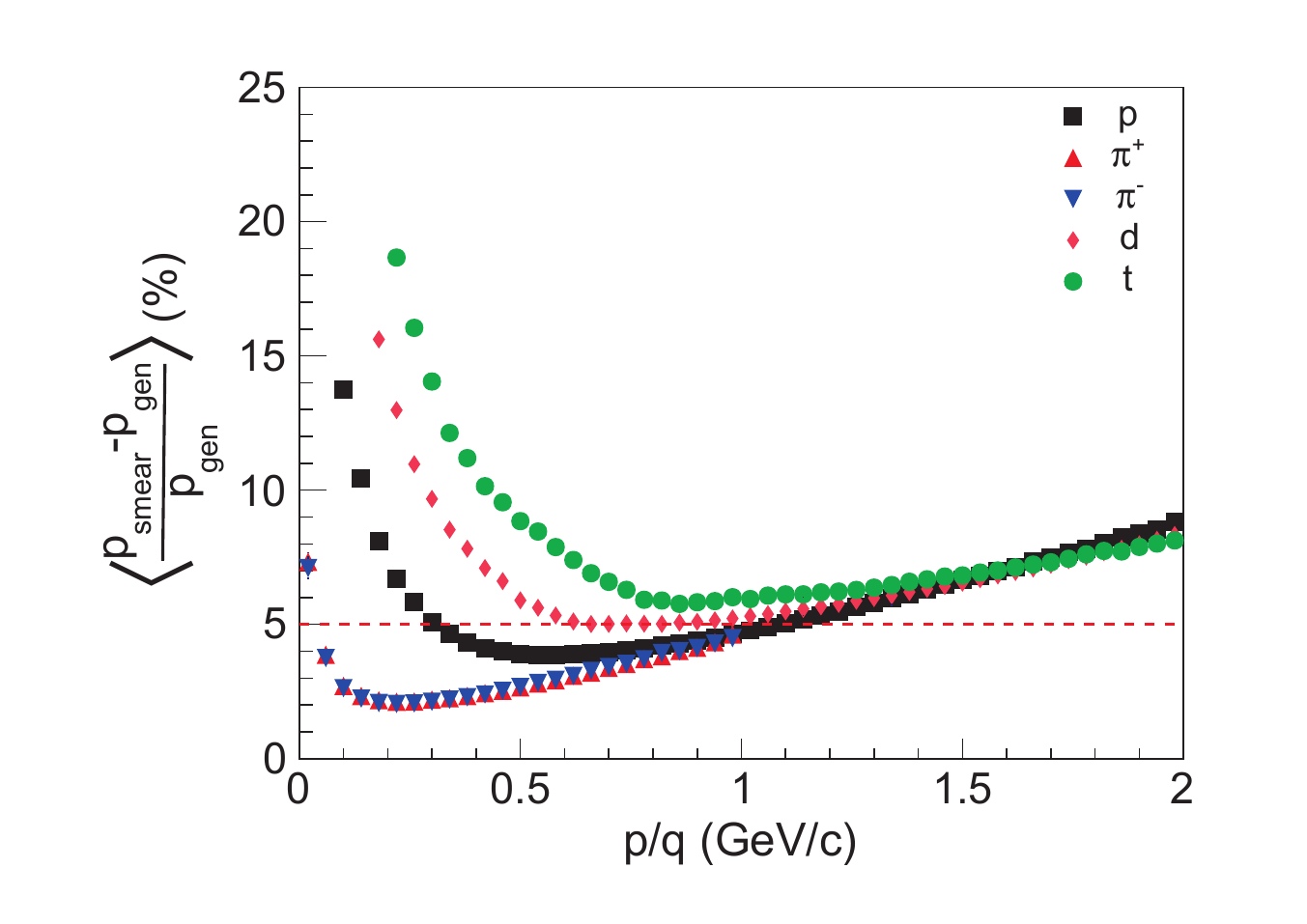}
\caption{(Color online) Momentum resolution of TPC in fast simulation}
\label{p-reso}
\end{figure}

 FIG.\ref{p-reso} shows the  resolution of the momentum reconstructed from the hit information collected from sensitive volume of TPC for  0.3 GeV/u $^{208}$Pb+$^{208}$Pb at $b=0$ fm.  $\sigma_{\rm xz}=0.6$ mm and  $\sigma_{\theta}=0.5 ^{\circ}$ are applied. 
 It is shown that  the typical resolution of $\pi^{\pm}$, proton and deuterium is about $5\%$.   With the increase of momentum or in the forward angle region, the momentum resolution is gradually degraded, suggesting the necessity of using  tracking detector array at forward angle for the dipole-type spectrometer. The degrades of the resolution at  low momenta is  mainly due to the loss of tracking efficiency because the tracks with low momentum leave very short length in the sensitive volume. 

\subsection{Global trigger signal construction}\label{sec. III3}

In the beam experiment of the CEE, the construction of the global physical trigger signal has a twofold goals: 1) One has to select the collisions on the target and suppress the background events on the beam path; 2) It can provide trigger signals for the events of interested  collision centrality. Similar scheme is usually adapted in the existing spectrometers, for instance, as by FOPI ~\cite{Kim2004Study} and SAMURAI ~\cite{Kobayashi2013SAMURAI}. 

\begin{figure}[htb]
\centering
\hspace{-1.0cm}\includegraphics[width=0.48\textwidth]{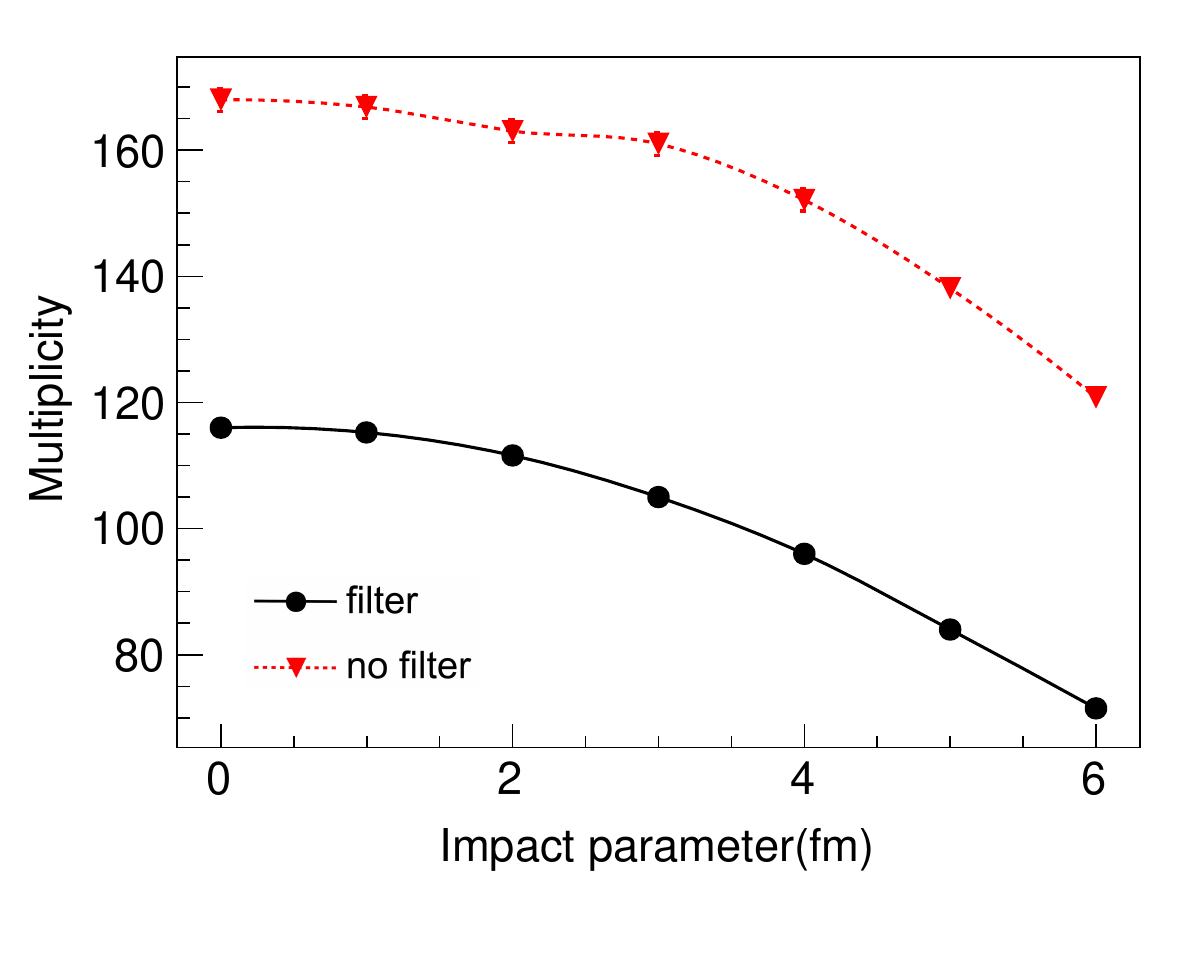}
\caption{(Color online) 0.5 GeV/u $^{208}$Pb+$^{208}$Pb reaction system, under the collision parameter b=0,1,2,3,4,5,6fm, the multiplicity of charged particles changes with the collision parameters before and after passing through the detector. The red (black) curve is the data before (after) the detector filtering.}
\label{mult-b}
\end{figure}

At CEE, the multiplicity of charged particles firing the fast detectors, including iTOF and eTOF, is employed to construct the global trigger because it can provide an approximate measure of the collision centrality. Figure \ref{mult-b} presents the  multiplicity of charged particles  as a function of the impact parameter $b$ in the transport model simulations of the reactions  $^{208}$Pb+$^{208}$Pb at 0.5 GeV/u. The multiplicity distributions  before (open) and after (solid) the filter of the iTOF and eTOF are presented.  It can be clearly seen that the multiplicity of  charged particles firing the iTOF and eTOF is anti correlated with $b$, giving certain ability to select the event centrality as the trigger signal of the CEE spectrometer.

\section{Study on the nEoS of nuclear matter} \label{sec. IV}

In this section, we present the feasibility of studying the nuclear equation of state using several selected observables at CEE. In the HIC in the energy domain of hundreds of MeV/u, the nuclear matter can be compressed to approximately $2\rho_0$, where $\rho_0$ is the saturation density. In addition, due to  the large stopping and large spatial-temporal volume, the sensitivity of the observable on nEOS is expected to be enhanced ~\cite{Zhang2009Systematic, Fu2008Nuclear}. Thus, it is of interest to check whether the information of nEOS carried by various the observables is discernible at CEE.  

\subsection{Production of light nuclei}\label{sec. IV1} 

One of the physical goal of CEE is to measure systematically the production of $^3\rm H$ and $^3 \rm He$ at the beam energies between 300 and 600 MeV/u. The motivation of this measurement is to understand the correlation between clustering and the isospin transport in HIC before arriving at a more stringent constraint on nuclear symmetry energy.  It has been proposed that the yield ratio of \rthe~ in HIC in wide energy range is a sensitive probe of nuclear symmetry energy~\cite{LieWen2003Effects, LieWen2004Effects, Qingfeng2005Probing, Zhang2005Probing,Yongjia2015ratio,Guo2012Influence,Hagel2000Light, LieWen2003Light}. However, the origin of $^3\rm H$ and $^3 \rm He$  are very complicated and the clustering process is reportedly correlated to the transport of isospin degree of freedom.  Experimentally, in the energy regime of 1 GeV/u, a systematic data of the production of pions, protons and light clusters including $^3\rm H$ and $^3 \rm He$ have been published by FOPI collaboration ~\cite{Reisdorf2010Systematics, Reisdorf2012Systematics}, but in some other experiments, the $\rm {^3H-^3He}$ puzzle has been reported ~\cite{Dempsey1996Isospin, Veselsky2001Isospin}.   

\begin{figure}[htb]
\centering
\hspace{-0.5cm}\includegraphics[width=0.45\textwidth]{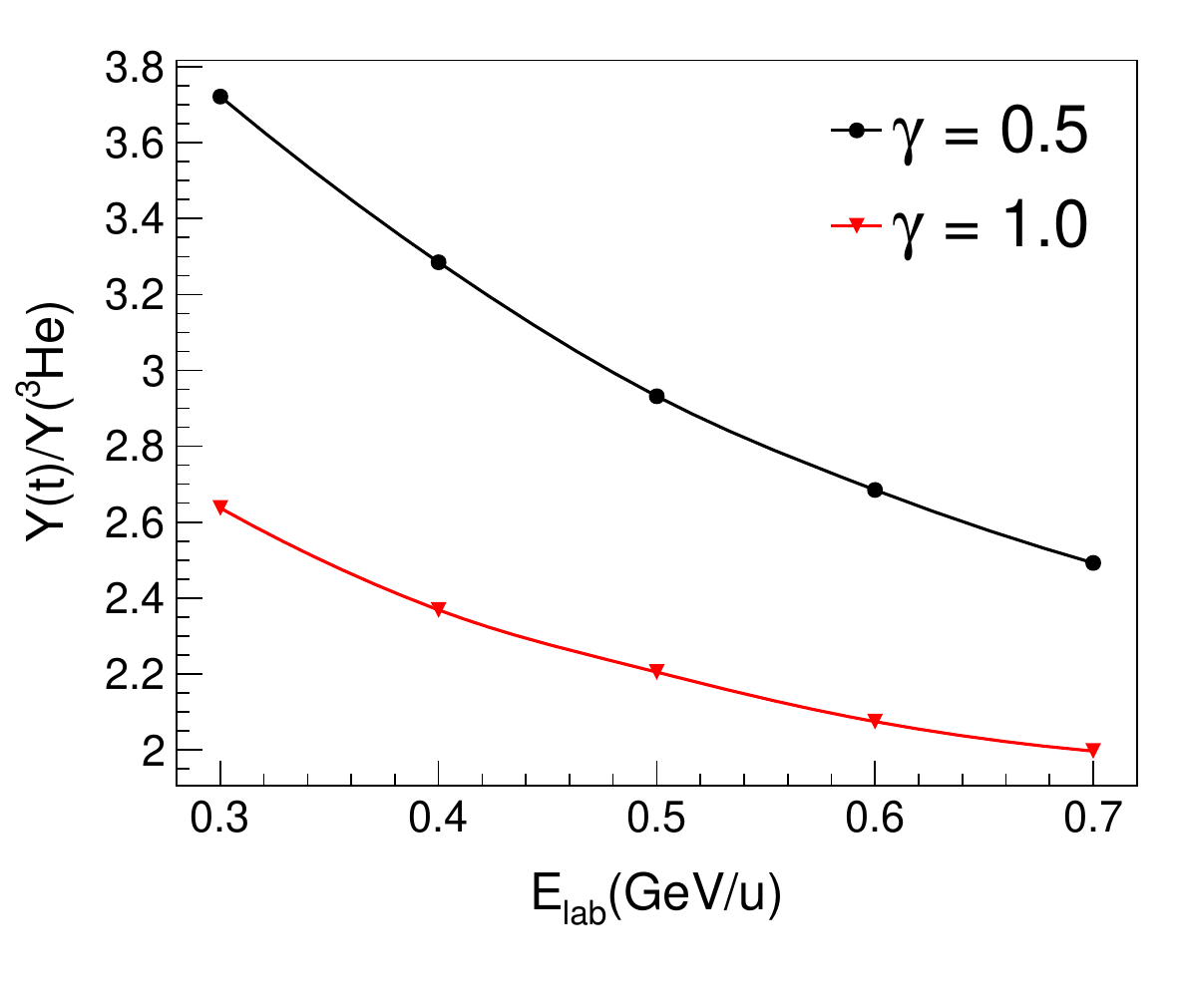}
\caption{(Color online) The $^{3}$H/$^{3}$He yield ratio as a function of beam energy for reaction system $^{208}$Pb+$^{208}$Pb at different symmetry energy parameters $\gamma=0.5$, 1.0}
\label{t/3He}
\end{figure}

With the aim of measuring the \rthe~ ratio, the events generated by UrQMD model  are analyzed in the fast simulation framework of CEEROOT. FIG.\ref{t/3He} presents the \rthe~ ratio as a function of beam energy of $^{208}$Pb+$^{208}$Pb system  given different slope parameter of \esym, $\gamma=0.5$ (soft) and 1 (stiff), respectively. The charged products are filtered by the tracking detectors. It is shown that the yield ratio \rthe~ exhibits certain sensitivity on \esym~ and the difference between $\gamma=0.5$ and 1 is kept after the detector filtering.

In order to clarify whether the symmetry energy takes effect on the clustering process, one can further investigate the degree of clustering by looking at the ratio of the number of free protons to that of bounded in clusters produced in the HICs. In a general picture, the isospin dependent N-N cross sections, the nuclear symmetry potential and the Coulomb interactions are convoluted in forming the clusters as well as the emission of free protons. Therefore, the yield  of the free protons, in comparison to the the bounded protons, carries presumably the information of the nuclear symmetry energy \esym, although the evolution of the system from a compression phase at early stage to the freeze-out stage is complicated and depends  on the transport models. 

\begin{figure}[htb]
\centering
\hspace{-0.5cm}\includegraphics[width=0.45\textwidth]{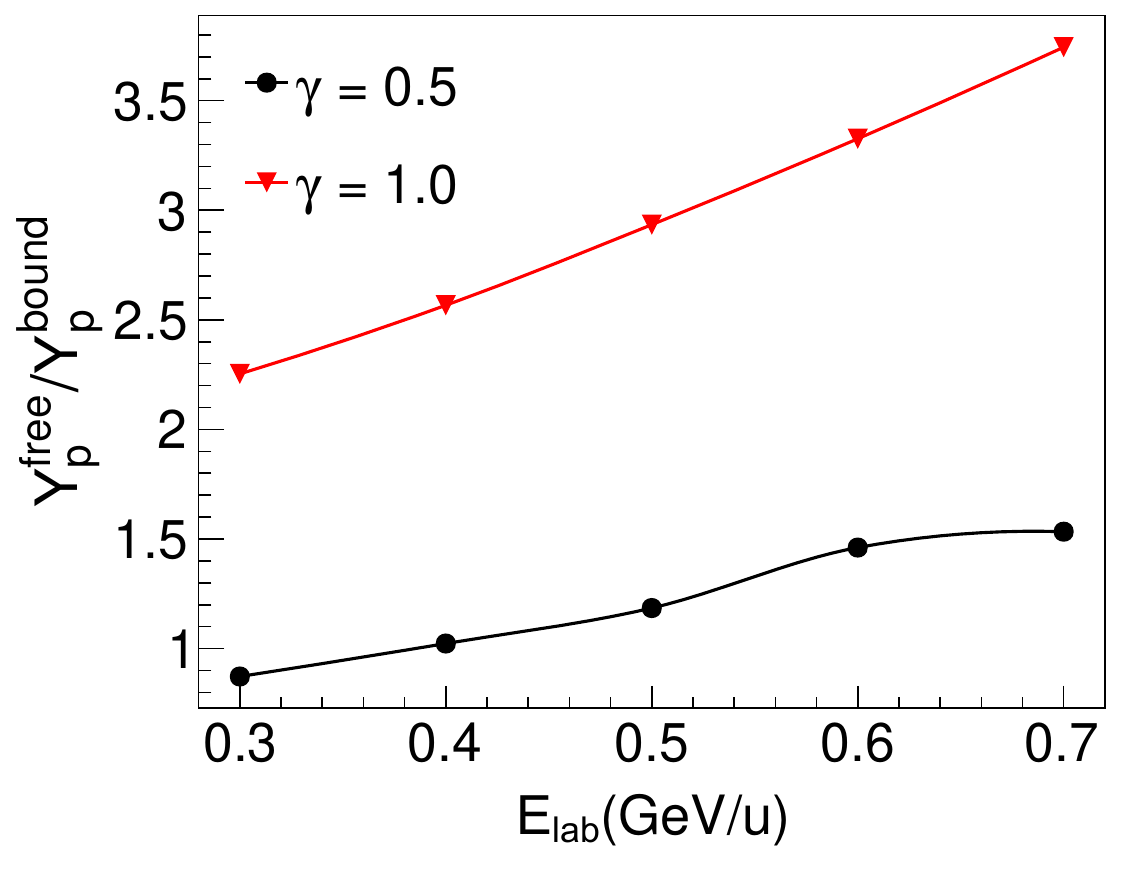}
\caption{(Color online) The ratio of free protons to that bounded in light clusters as a function of beam energy in  $^{208}$Pb+$^{208}$Pb reactions. Two parameters of \esym, $\gamma=0.5$ and 1.0 are compared.}
\label{bound-p}
\end{figure}

FIG.\ref{bound-p} presents the excitation function of the ratio of the free and bounded protons for the central collisions  $^{208}$Pb+$^{208}$Pb at 300 MeV/u using different $\gamma =0.5$ and 1.0, respectively. The yield of bounded proton counts the charge of the light particles with $Z<3$ in the final state of the reaction, including d, t, $^{3}$He and $^{4}$He. The  ratio increases gradually with the  beam energy, indicating that the clustering is more favored at low beam energy. Moreover, it is shown that the degree of clustering, measured by the ratio $Y_{p}^{\rm free}/Y_{p}^{\rm bound}$, can differentiate the stiffness of \esym. With  a softer symmetry energy ($\gamma=0.5$), more protons are found in the clusters. 

\subsection{Radial flow and nEOS}\label{sec. IV2} 

It has been realized since long time that the collective flow is an observable of nEOS under extreme conditions ~\cite{Ollitrault1998Flow, Bauer1993Large, Poggi1995Evidence, Reisdorf2004Nuclear, Stoicea2004Azimuthal, Stachel1996thermalization, Bao1996Pion}. Depending on the collision geometry, the collective flow has different forms. In non central collisions, collective motion of HIC manifests itself in directional flow and elliptic flow, denoting the first and second order coefficient of the Fourier expansion of the azimuthal distribution of the final products, respectively ~\cite{Ollitrault1993Determination}. In central collisions,  the fireball produced in the collision experiences compression following by an expansion, forming the so-called radial flow  ~\cite{Alard1992Midrapidity}. It can be expected that the size of radial flow depends on the compressibility of nuclear matter ~\cite{Hartnack2001Transverse, LieWen2003Light}. Later, it is demonstrated by FOPI collaboration that the radial flow is sensitive to the incompressibility coefficient of nuclear matter ~\cite{Reisdorf2010Systematics}. In this subsection,  the detection of radial flow with CEE is discussed. 

The radial flow can be  extracted from the transverse momentum spectra of the final particles at midrapidity in the central collisions. Two methods have been developed to obtain the expansion velocity $\beta_r$, one is to fit the $p_{\rm t}$ spectra using  Siemens-Rasmussen formula ~\cite{Siemens1979Evidence}, the other is to extract the dependence of the mean kinetic energy as a function of the mass of final products based on blast wave model ~\cite{Gustafsson1984Collective, Lisa1995Radial, Wang1996Retardation}. In both methods, the thermal motion characterized by the temperature $T$ and the expansion velocity $\beta_r$ are included.  Here it is attempted to study the radial flow by the Siemens-Rasmussen formula ~\cite{Siemens1979Evidence} written as:

\begin{equation}\label{eqn-3} 
    \frac{1}{p_{\rm t}}\frac{d^{2}N}{dp_{\rm t}dy^{0}}=CEe^{-\gamma E/T}[(\gamma+\frac{T}{E})\frac{\sinh\alpha}{\alpha}-\frac{T}{E}\cosh\alpha]
\end{equation}
where $\gamma=1/\sqrt{1-\beta_{r}^{2}}$, $\alpha=\beta_{r}p\gamma/T$, $E$ and $p$ are the energy and momentum of the particle in the center of mass reference. The radial expansion velocity $\beta_{r}$, thermal freeze out temperature $T$ and the  constant $C$ are the fitting parameters. 

\begin{figure}[htb]
	\centering 
	\subfigure{
        \includegraphics[width=0.49\linewidth]{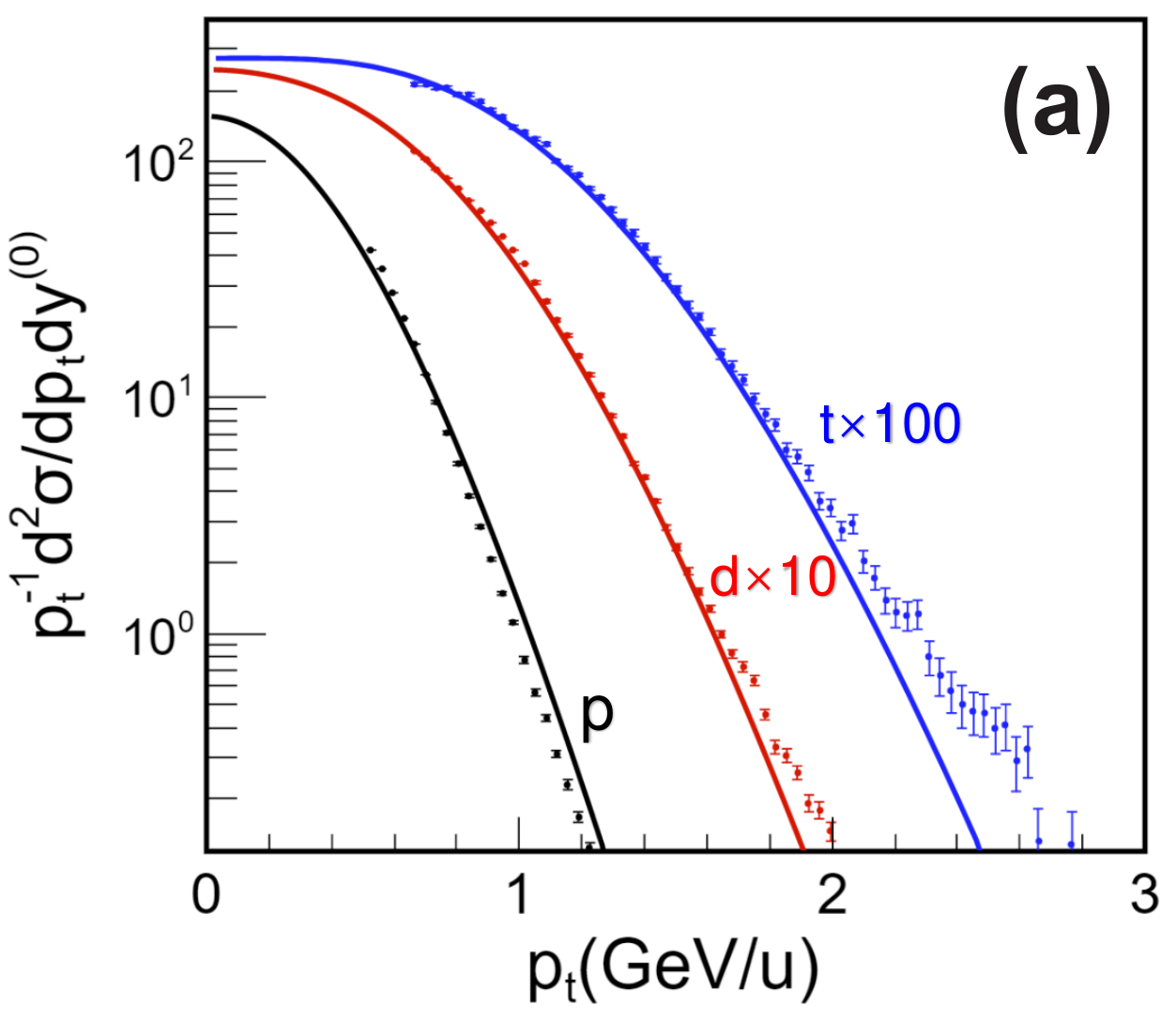}}
	\subfigure{
		\includegraphics[width=0.47\linewidth]{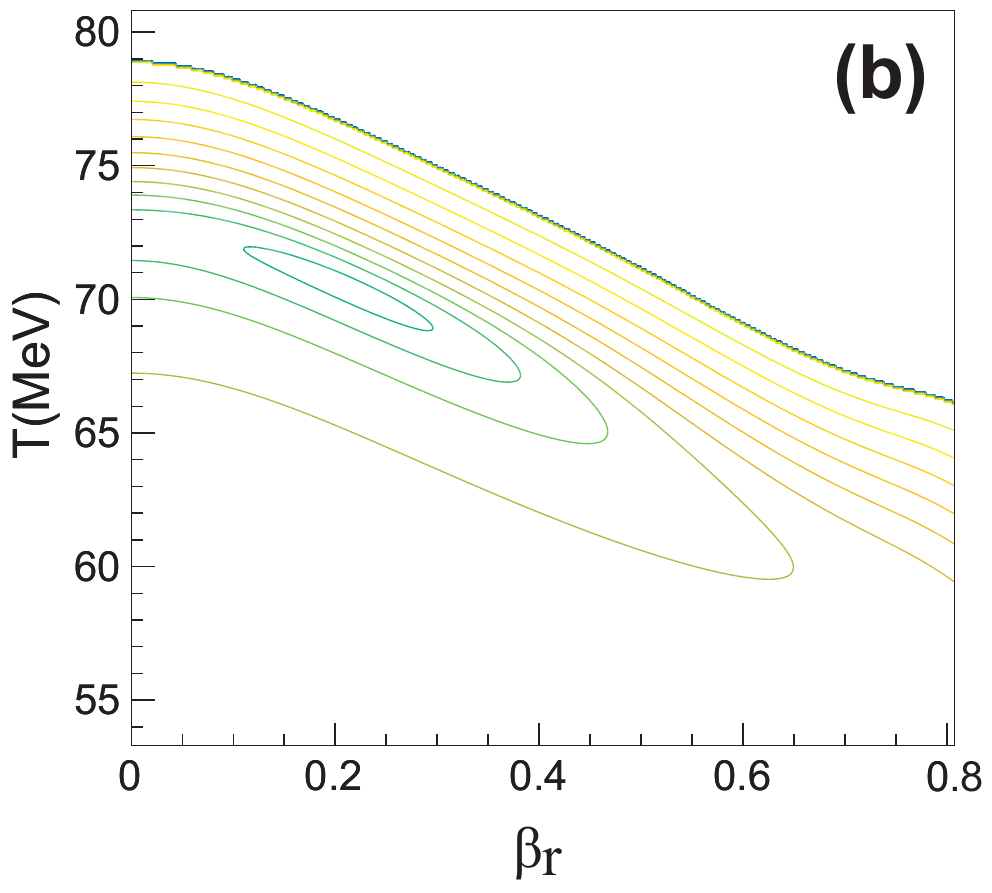}}
	\caption{(Color online) (a) The $\beta_{r}$ and T are the fitting curves of the $p_{\rm t}$ spectrum and the model in $^{208}$Pb+$^{208}$Pb at 0.4 GeV/u. The low transverse momentum part is eliminated in the fitting; (b) shows that $\chi^{2}/{\rm NDF}$ is fitted as $\beta_{r}$ and T.}
	\label{chi2 and pt_fit} 
\end{figure}

FIG.\ref{chi2 and pt_fit} presents the feasibility of the radial flow measurement in central collisions of $^{208}$Pb+$^{208}$Pb  at CEE.  Panel (a) presents the $p_{\rm t}$ spectra of proton, deuteron and triton. The low momentum parts are filtered by the acceptance cut and track length cut.  The curves are the simultaneous fitting results using formula (\ref{eqn-3}). It can be seen that the $p_{\rm t}$ spectra can be reproduced fairly well by Siemens-Rasmussen formula. In panel (b), the minimum $\chi^2$ distribution is plotted on the $T-\beta_{\rm r}$ plane. It can be seen that by simultaneous fitting, the two parameters can be determined rather well. While in individual fitting to the $Z=1$ isotopes, the two parameters are correlated.

\begin{figure}[!htbp]
	\centering 
	\subfigure{
        \hspace{-0.5cm}\includegraphics[width=0.51\linewidth]{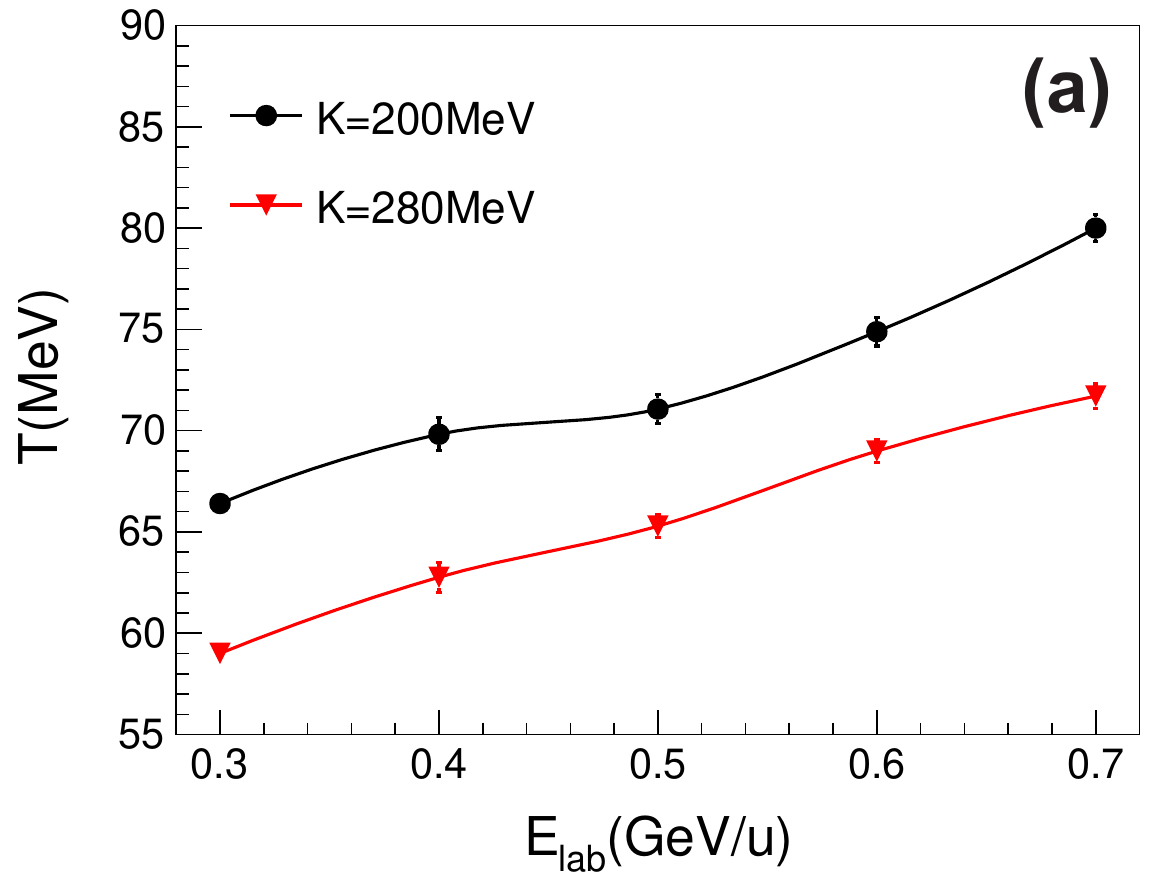}}
	\subfigure{
		\includegraphics[width=0.49\linewidth]{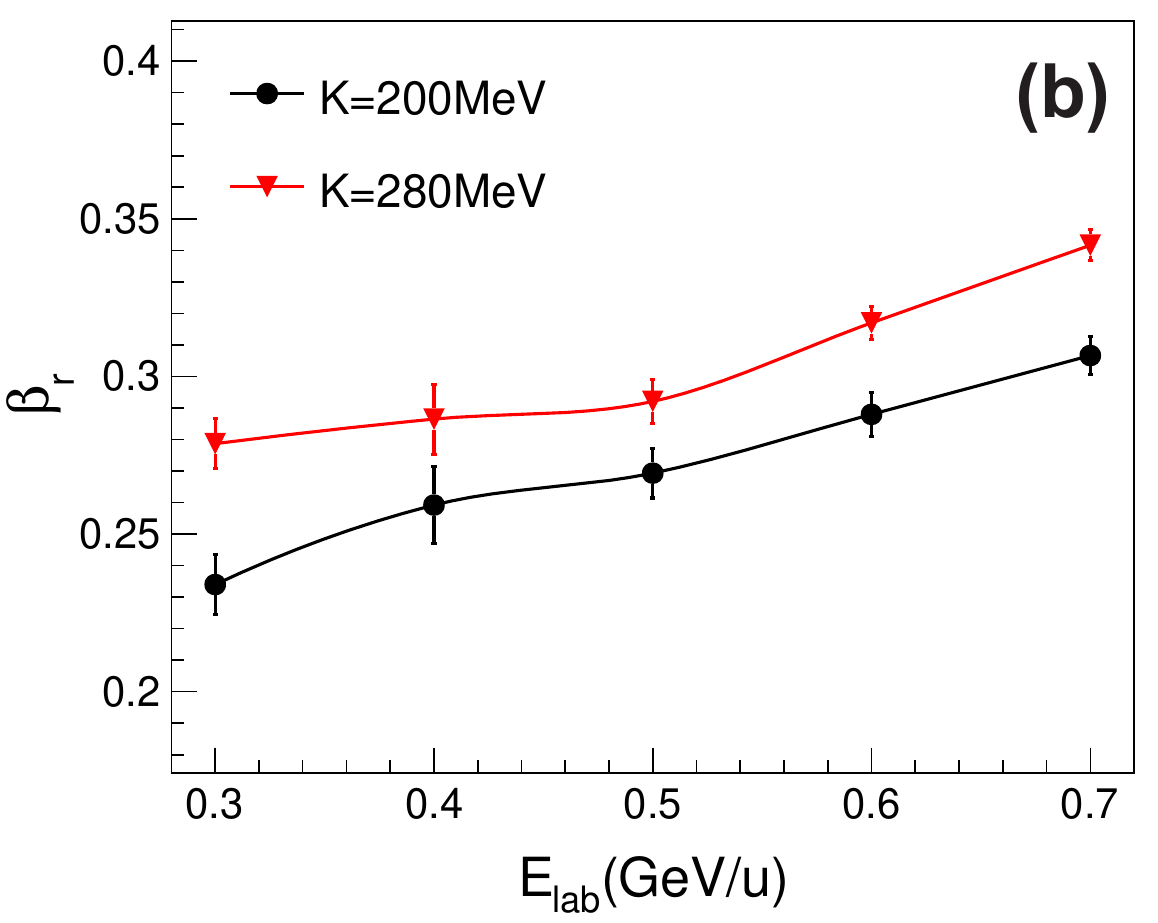}}
	\caption{(Color online) (a) It is shown that the thermal freeze out temperature T increases monotonically with the increase of beam energy in the energy range of 0.3 GeV/u to 0.7 GeV/u. (b) Shows radial flow $\beta_{r}$ increases monotonically with the increase of beam energy in the energy range from 0.3 GeV/u to 0.7 GeV/u.}
	\label{beta T} 
\end{figure}

FIG.\ref{beta T} presents the  excitation function of $\beta_{r}$ (a)  and $T$  (b) reconstructed from the simulations  with $\kappa=200$ and 280 MeV, repsectively. The transport model simulation reveals that different incompressibility leads to different freeze-out temperature $T$ and expansion velocity $\beta_{\rm r}$. With a soft  incompressibility constant, $T$ is  higher and $\beta_r$ is lower, relatively, and {\it vise versa}. The difference is larger than $10\%$ for most of the cases. It suggests that hard nuclear matter causes violent compression and expansion, while the soft nuclear matter tends to convert more kinetic energy to thermal energy in the procedure of the collision. It is also suggested that CEE has the ability for the studies of radial flow in the collisions of heavy system. 

\subsection{$\pi^{-}/\pi^{+}$ yield ratio and symmetry energy \esym}\label{sec. IV3}

The yield ratio of  $\pi^{-}/\pi^{+}$ produced in HIC has been proposed as a sensitive probe of nuclear symmetry energy at supra-saturation densities~\cite{Xiao2014Probing, Bao2005Near, Hong2005Charged, Qingfeng2005Probing, Cozma2011Neutron}. In a transport picture, the pions are  mainly produced by $\Delta$ resonance ~\cite{Yangyang2021Insights,Stock1986Particle}. From the isobaric model, the initial ratio $(\pi^{-}/\pi^{+})_{\rm res}$=$(5N^{2}+NZ)/(5Z^{2}+NZ)\approx(N/Z)^{2}_{\rm dense}$, where N and Z are the number of neutrons and protons in the reaction participating region. Therefore, $(\pi^{-}/\pi^{+})_{\rm res}$ is a direct measure of the isospin asymmetry $(N/Z)_{\rm dense}$, wihch is relatively higher than that of the system  due to the isospin fractionation effect arising from $E_{\rm sym}(\rho)$. In return, the yield ratio  $\pi^{-}/\pi^{+}$ becomes a sensitive probe of  $E_{\rm sym}(\rho)$  ~\cite{Bao2005Near}. In a statistic model ~\cite{Bonasera1987Isospin}, the ratio $\pi^{-}/\pi^{+}$ is related to the chemical potential of neutron and protons written as $\pi^{-}/\pi^{+}\sim \exp[2(\mu_{n}-\mu_{p})/T]$, where $T$ is the system temperature, and $\mu_{n}-\mu_{p}$ is the difference between the chemical potentials of neutrons and protons, which depends on  $E_{\rm sym}(\rho)$ ~\cite{Bao2002Probing}.  Circumstantial evidences of a soft \esym~ has been found from the pion data of FOPI collaboration in accordance with the isospin fractionation picture ~\cite {Xu2000Isospin}. However, the conclusion turns to be inconclusive because of the model dependence and the observable dependence, since the modeling of the pion production transport, as well as clustering, is very complicated in transport model. For this direction, one can refer to ~\cite{Reisdorf2007Systematics, Xiao2009Circumstantial, XIAO2011Isospin}.  

\begin{figure}[!htbp]
\centering
\hspace{-0.5cm}\includegraphics[width=0.45\textwidth]{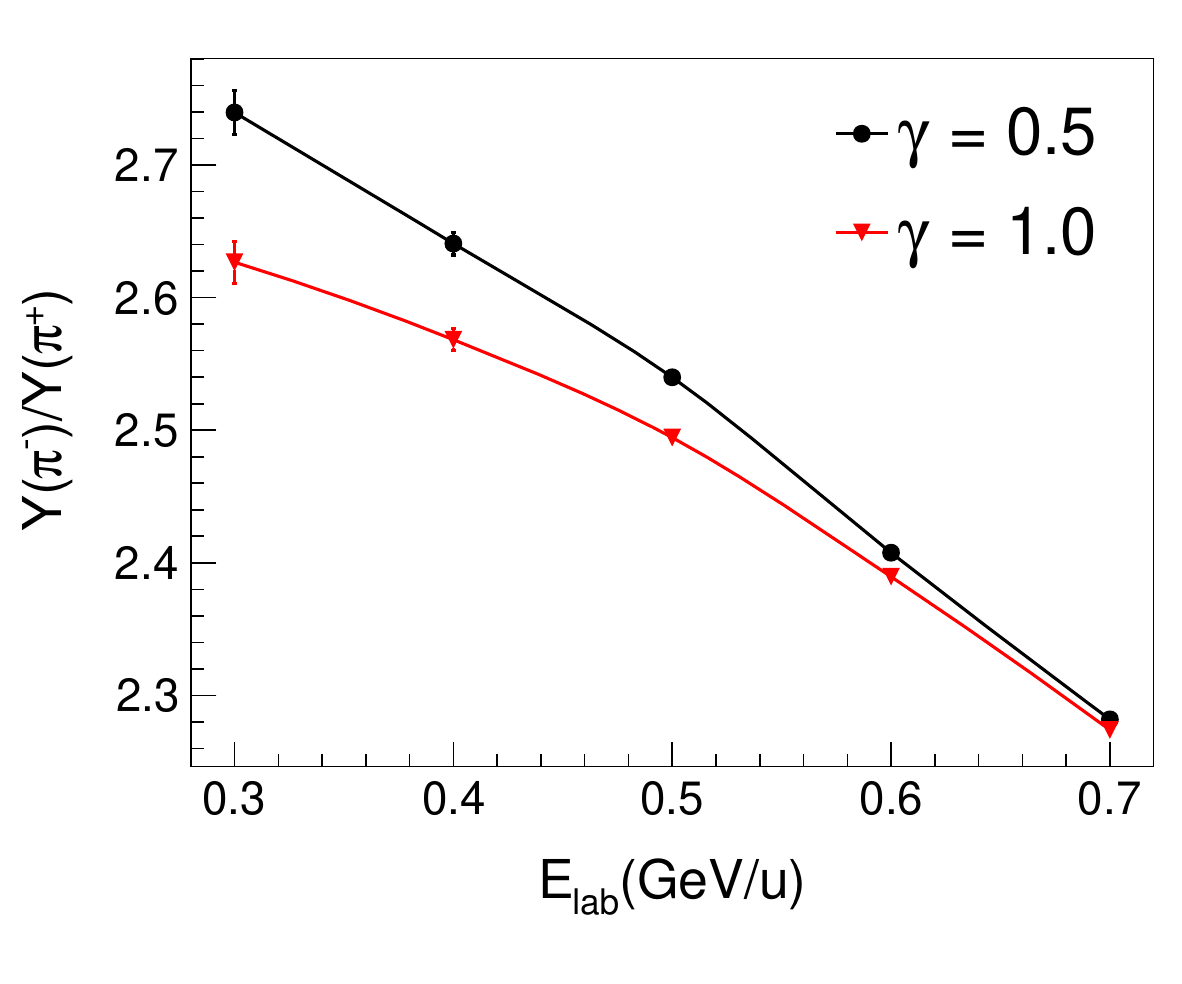}
\caption{(Color online) The  excitation function of final $\pi^{-}/\pi^{+}$ yield ratio in $^{208}$Pb+$^{208}$Pb reactions with different symmetry energy parameters $\gamma=0.5$ and 1.0, respectively.}
\label{pion}
\end{figure} 

Therefore, pion physics is one of the focuses at CEE. In the current design of CEE, as shown in  FIG.\ref{PbPb phase space}(f),(g) in section \ref{sec. III1}, the whole phase space of pions  on $u_{\rm t}^{(0)}-y^{(0)}$ plane is well covered, although the efficiency on azimuth is lost near $\phi=90^\circ$ and $270^\circ$ in laboratory. With the current efficiency, the final yield of $\pi^{-}/\pi^{+}$ in $^{208}$Pb+$^{208}$Pb reactions at various  beam energies are extracted and plotted in FIG.\ref{pion} with $\gamma=0.5$ and 1.0, respectively.  It is shown that the ratio $\pi^{-}/\pi^{+}$ decreases with increasing of the beam energy. The difference between $\gamma=0.5$ and 1 is also more pronounced at lower beam energy, in accordance with previous predictions for Au+Au systems ~\cite{Xiao2009Circumstantial}. This trend is due to the larger space-time volume of the fireball created in the collisions at lower energies ~\cite{Zhang2009Systematic}. It is suggested that at the favorable energy region of CEE, namely near 300 MeV/u , the stiffness of \esym~ between $\gamma=0.5$ and 1.0 can be discriminated   if the systematic error can be controlled within $10\%$.

\subsection{$K^{0}_{S}$ and incompressibility}\label{sec. IV4}

Finally, let us discuss briefly the production and detection of kaons at CEE. Kaons are mainly produced in the supra-saturation density zone created in the early stage of the collisions, and the typical production channels are $B+B\rightarrow B+Y+K$, $\pi+B\rightarrow Y+K$. Kaon production is of interest in this energy region as a very clean probe to nEOS because it experiences much less rescattering process in comparison to that of nucleons or pions. Since long time it has been pointed out that the yield of kaons is an effective probe for nEOS of supra-saturation density nuclear matter ~\cite{Aichelin1985Subthreshold}. KaoS collaboration from GSI  measured the yield of $\rm K^+$  in C+C and Au+Au systems from 800 MeV/u to 1.6 GeV/u. Within the transport model framework,  the yield ratio of $\rm K^+$ in the two systems favors a soft nuclear matter ~\cite{Sturm2001Evidence, Hartnack2006Hadronic}.  Compared to  charged Kaons, the neutral kaon $\rm K^0$ is not affected by Coulomb interactions, so the production of $\rm K^0$ carries the information of the high-density medium where it is produced.   The yields of these neutral strange particles were studied in detail on the SIS, BEVALAC and AGS ~\cite{Du:2018ruo,Crochet2000Sideward,Albergo2002Spectra}.

\begin{figure}[htb]
\centering
\hspace{-0.6cm}\includegraphics[width=0.5\textwidth]{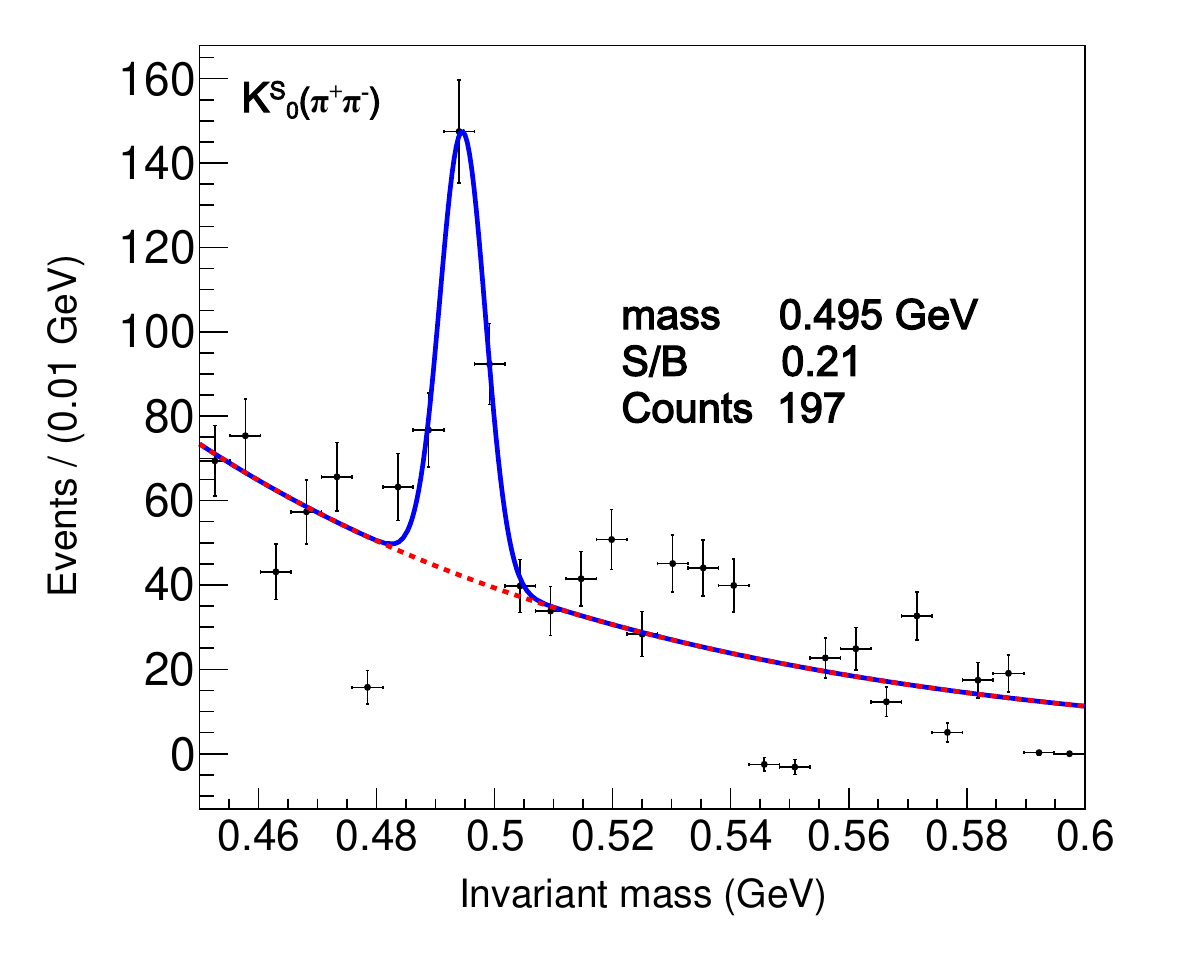}
\caption{(Color online) $K^{0}_{S}$ invariant mass spectrum of p+Ni system at K=200 MeV}
\label{mass inv}
\end{figure}

Since the HIRFL-CSR can provide proton beam of the maximum energy 2.8 GeV and carbon beam of the maximum energy 1.2 GeV, it is also possible to conduct kaon experiments at CEE. Since the identification of  $\rm K^+$ may suffer from the heavy background caused by protons and pions, it is of our interest as the first step to measure the neutral $\rm K^0$ decaying to $\pi^+$ and $\pi^-$, which are thus easier to be detected.  FIG.\ref{mass inv} shows the invariant mass spectrum of $\pi^+$ and  $\pi^-$ in p+Ni system at $\kappa=200$ MeV. It is shown that, after the background are subtracted by mixing event method, a peak at $K^0$ mass is evidently seen.    Totally 1 M central collision events are simulated and about 200 $\rm K^0_{\rm s}$ are reconstructed. Worth mentioning, the background is not fully removed and there is an enhancement at low mass end, this is not due to the detector filtering, but possibly due to the residue baryon-baryon correlation since the majority of pions are produced via $\Delta$ resonances.  Compared to the Monte-Carlo truth tracks, the reconstruction efficiency of $K^0$ is about $35\%$.

\begin{figure}[htb]
\centering
\hspace{-0.7cm}\includegraphics[width=0.46\textwidth]{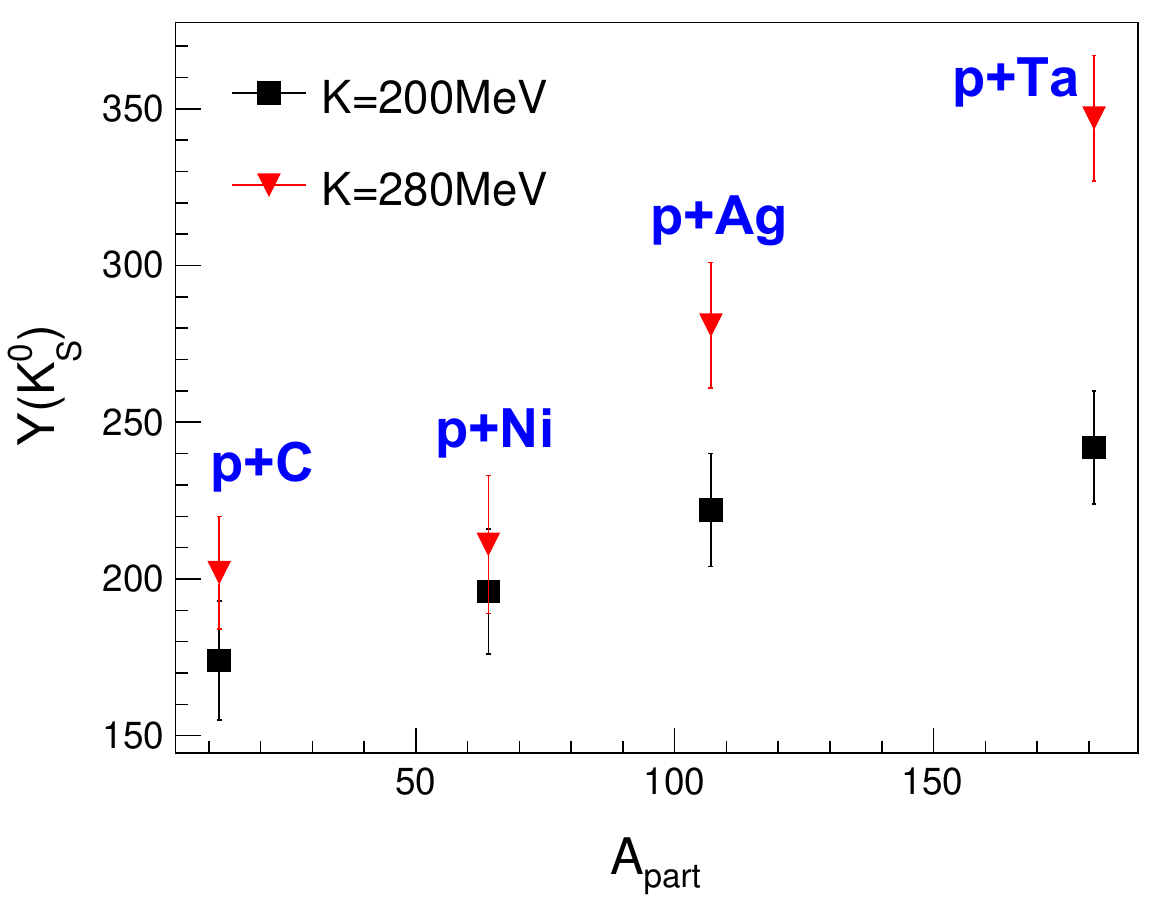}
\caption{(Color online) Variation of $K^{0}_{S}$ yield in p+C, p+Ni, p+Ag and p+Ta reactions at beam energy of 2.8 GeV with two incompressibility $\kappa=200$ and 280 MeV, respectively.}
\label{incomps}
\end{figure}

FIG.\ref{incomps} shows the yield of the reconstructed $K^{0}$ (mainly $K^{0}_{\rm s}$) as a function of system size with two  different incompressibility  coefficients $\kappa=200$ and 280 MeV, respectively.  The yield of $K^{0}$ increases with the mass of the system, and exhibits significant dependence on the incompressibility $\kappa$ of the nuclear matter. As long as the systematic uncertainty can be controlled within $10\%$, the production of neutral Kaon $K^0$ can also be used as a probe of the nuclear equation of state, supporting one of the physical programs of CEE.

\section{Conclusion} \label{sec. V}

The conceptual design  of CEE are briefly described. The expected performance is studied  based on the Geant 4 simulations using UrQMD as the event generator. The main part of the detector system is a large-acceptance magnetic dipole housing the tracking detectors and the surrounding TOF detecotrs.  TPC and MWDC are employed to cover the midrapidity and forward rapidity region for tracking, respectively. TOF detectors, including iTOF and eTOF, provide the main trigger signals constructed based on advanced FPGA technology. In the framework of CEEROOT, built {\it ad hoc} for the researches and developments ($R\&D$) of CEE, we have discussed the feasibility of the studies of nuclear equation of state through various observables, including the production of light clusters, \rthe~ yield ratio, radial flow, pion ratio and near-threshold Kaon production. The results of the simulations suggest that the aforementioned probes can be measured with CEE, giving the opportunity to the studies of properties of dense nuclear matter in the HIRFL-CSR energy regime.

\section*{Declaration of competing interest}
The authors declare that they have no known competing financial interests or personal relationships that could have appeared to influence the work reported in this paper.

\section*{Acknowledgments}
 This work is supported  by the National Natural Science Foundation of China under Grant Nos. 11927901 and 11890712  
  and by Tsinghua University Initiative Scientific Research Program.

\bibliographystyle{unsrt}
\bibliography{reference}

\end{document}